\documentclass[]{revtex4-2}
\usepackage{amsfonts,amssymb,amsmath}
\usepackage{graphicx}
\usepackage{multirow}
\usepackage[pdfa]{hyperref}
\usepackage[all]{hypcap}

\hypersetup{
    colorlinks=true,
    linktoc=all,
    menucolor=black,
    citecolor=blue,
    urlcolor=blue
}

\hypersetup{
    linkcolor=black
}

\hypersetup{
    linkcolor=blue
}

\graphicspath{{fig/}}

\begin{document}

\title{Beamstrahlung monitoring at SuperKEKB upgrade 2023.}

\author{D.~Liventsev\,\footnote{\href{mailto:livent@wayne.edu}{\texttt{e-mail:livent@wayne.edu}}}}
\author{G.~Bonvicini\,\footnote{\href{mailto:gbonvicini@wayne.edu}{\texttt{e-mail:gbonvicini@wayne.edu}}}}
\author{D.~Ricalde~Herrmann}
\affiliation{Wayne State University, Detroit, Michigan 48202, U.S.A.}

\author{P.~L.~M.~Podesta-Lerma}
\affiliation{Universidad Autonoma de Sinaloa, Sinaloa 80000, Mexico}

\author{M.~Tobiyama}
\affiliation{High Energy Accelerator Research Organization (KEK), Tsukuba 305-0801, Japan}

\begin{abstract}
  Beam monitoring is crucial for particle accelerators to achieve high
  luminosity. We describe an upgrade of the LABM, a beam monitoring
  device utilizing observation of the beamstrahlung, radiation emitted
  by a beam of charged particles when it accelerates in the
  electromagnetic field of another beam of charged particles.
\end{abstract}

\maketitle


\section{Introduction}

The main task of the modern particle physics is a search for the
phenomena beyond the Standard Model (BSM). Since BSM effects are
expected to be small, we need large data samples, which requires stable
accelerator operation with high luminosity. To achieve it beam
monitoring is essential.

SuperKEKB~\cite{skekb1,skekb2} is an asymmetric-energy $e^+e^-$ collider
particle accelerator with a circumference of 3.016$\,$km that provides
luminosity for the Belle~II experiment~\cite{belle2}. SuperKEKB is the
successor of KEKB~\cite{kekb}. While KEKB achieved a maximum peak
luminosity of $2.11 \times 10^{34}\,\textrm{cm}^{-2}
s^{-1}$~\cite{kekb}, SuperKEKB aims to reach a peak luminosity of $8
\times 10^{35}\,\textrm{cm}^{-2} s^{-1}$~\cite{skekb1}, about 40~times
larger than its predecessor. Since the luminosity is strongly dependent
on beam-optical parameters at the IP, to have direct measurement of such
parameters is a key factor for success. For a nano-beam collider such as
SuperKEKB, where there is strong sensitivity to small parameter changes,
such information is especially important in order to maximize the
luminosity yielded at a bunch crossing.


\section{Large angle beamstrahlung monitor}

The device that we describe in this paper is the Large angle
beamstrahlung monitor (LABM). Beamstrahlung is the radiation emitted by
a beam of charged particles when it accelerates in the electromagnetic
field of another beam of charged particles at the interaction point
(IP)~\cite{bs}. Beamstrahlung polarization and spectra are closely
related to the beam configuration at the IP, and therefore the LABM can
be used to passively monitor the position and size of the beams directly
at the IP. The beamstrahlung was first observed at SLC~\cite{labm-slc},
the linear electron-positron collider at SLAC. The first LABM prototype
was successfully tested at CESR~\cite{labm-cesr}, a $e^+ e^-$ collider
located at Cornell University, and operated between 2007 and 2008. An
upgraded version of the LABM was installed at SuperKEKB in
2015~\cite{labm-skekb-1} and operates since 2016~\cite{labm-skekb-2}.


LABM operating at SuperKEKB accelerator consists of four telescopes
pointed at the Belle~II IP, at around 8.5~mrad observation angle from
the beam axis. On each side of Belle~II detector there are two
telescopes on the top (azimuthal position of 90 degrees) and bottom
(270 degrees) of one of the beam pipes. This redundancy improves the
systematics and allows to be sensitive to vertical asymmetries of the
particle beams.

\begin{figure}[htp]
  \includegraphics[width=0.5\textwidth]{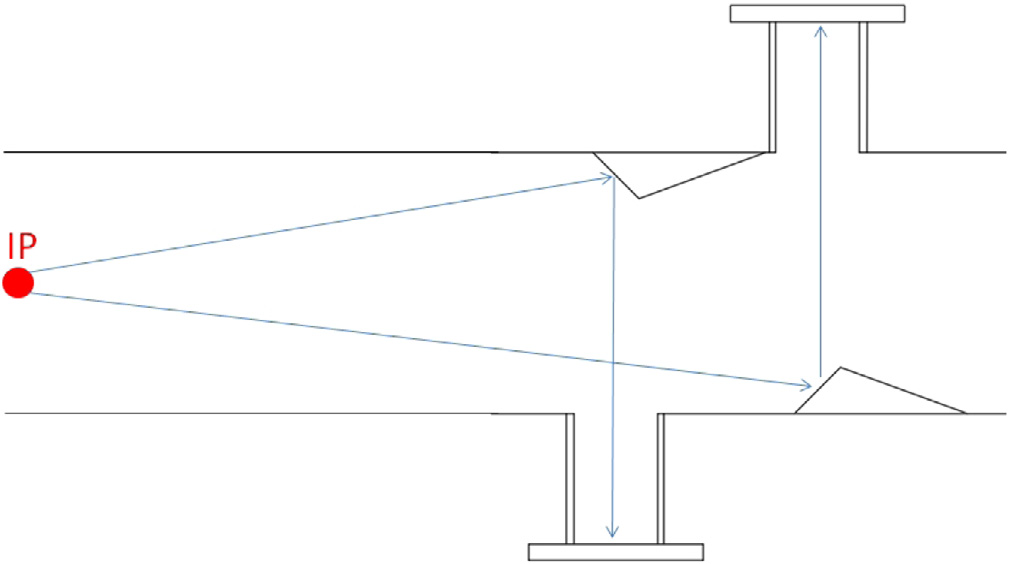}\\
  \caption{Two vacuum mirrors are used to extract the beamstrahlung
    light from the vacuum beam pipe through small windows. The
    beamstrahlung light coming from the IP is indicated with
    arrows. The light is then guided towards the detector by a system
    of pipes and mirrors, which is not shown here.}
  \label{p:labm-mirrors-beampipe}
\end{figure}

Beamstrahlung is intercepted inside the beam pipe with vacuum mirrors
made of beryllium at about 5~meters downstream from the IP and extracted
through a special window, as shown in
Fig.~\ref{p:labm-mirrors-beampipe}. The size of the mirrors is $2 \times
2.8\,\mathrm{mm}^2$, which corresponds to a square angular acceptance
once the mirror inclination is taken into account. Then the light
travels inside a series of straight aluminum pipes with mirrors at pipe
joints until it arrives to an optical box located outside of the
interaction region. There are two optical boxes, one for the electron
beam (``Oho'' side of the detector) and one for the positron beam
(``Nikko'' side of the detector). The first mirror after the vacuum
window, referred to as a ``primary mirror'', may be moved with the help
of two stepper motors and may be controlled remotely, which allows to
perform an angle scan in search of the signal.

In the optical boxes the light is split into two transverse polarization
components by Wollaston prisms. The beamstrahlung is expected to be
highly polarized, so that we can observe the signal in one polarization
and use the other for the background estimation. In the classical LABM
design used until 2023 each component was observed at four different
average wavelengths by photomultiplier tubes (PMT). See
Refs.~\cite{labm-skekb-1} and~\cite{labm-skekb-2} for more details of
that design. After an upgrade in 2023, which is described in this paper,
the PMTs were replaced with cameras.



After several years of studies and upgrades in 2022 we could
successfully demonstrate that the LABM measurements of beamstrahlung
signal are of good quality and are correlated with key beam
parameters. Due to the problems with getting the proper IP spot for the
positron beam, only two telescopes monitoring the electron beam were
used, \textit{i.e.} measurements from 16~PMTs. Data sample was collected
with telescopes looking at the IP without moving for 11~days. Only
Physics run data with currents above $100\,$mA was used. The results
were obtained with a Machine Learning (ML) using a Neural Network (NN).
External measurements of the beams sizes from the accelerator group and
luminosity from the detector group were used for training. The mean
absolute error (MAE) of prediction of the beam parameters on a data
sample not used in the training was at a few percent
level~\cite{nnresult}.

\section{The 2023 upgrade}

The classic LABM design had several drawbacks. First, PMTs provide
only a point-like measurement. To get an image of the IP including
areas surrounding the spot a scan was necessary which could take
significant time of the order of hours, while beam conditions change
every second. It was possible to take measurements in one point
without moving but it provided a limited information. Besides, over
longer periods the spot might move and tracking it needed a new scan.

Second, since the optical boxes are located outside of the beamline, it
is necessary to use long optic paths with many mirrors which need
adjustment. The adjustment procedure is error prone and there is always
a risk of choosing a reflection, a fake spot etc instead of a true spot.


In 2023 we performed a major LABM upgrade and replaced PMTs with CMOS
a2A1920-51gcPRO cameras from Basler~\cite{basler}. This allowed us to
capture the image of the IP spot and backgrounds together.  Because of
this and since the cameras make an image (measurement) in three ranges
of the spectrum at once, \textit{i.e.} in red, green and blue, most of
the internals of the optic boxes became unnecessary.

\begin{figure}[htp]
  \includegraphics[width=0.3\textwidth]{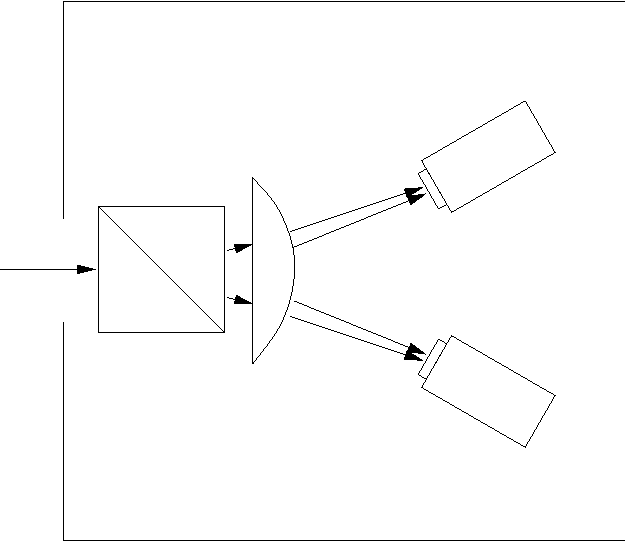}
  \put(-115, 100){1}
  \put( -90, 100){2}
  \put( -40, 105){3}
  \put( -40,  53){3}
  \caption{Optic box with cameras. The light enters the box as indicated
    by an arrow, the Wollaston prism (1) splits the light polarizations,
    the lens (2) focuses the light on the cameras (3).}
  \label{p:labm-camera-scheme}
\end{figure}

We expect the spot to have a size of approx. $4 \times
4\,\textrm{mm}^2$. We would also like to have an image of the area
around the spot to estimate backgrounds. Since the size of the sensor is
$6.6 \times 4.1\,\textrm{mm}^2$, we use a focusing lens after the
Wollaston prism. The focal length of the lens is $40\,$cm, and the
cameras are installed at $30\,$cm from the prism, as shown in
Fig.~\ref{p:labm-camera-scheme}. The exposure time of cameras was set to
$0.1\,\mathrm{s}$, no additional lens used. Gratings, mirrors, PMTs,
electronics and conveyor belts were removed from the optical boxes.

We also replaced primary mirrors drives: stepper motors, controllers and
cables to avoid interference problems which we observed before. The
optic path remained the same for this data taking period.

\begin{figure}[htp]
  \begin{tabular}{ccc}
    \includegraphics[width=0.48\textwidth]{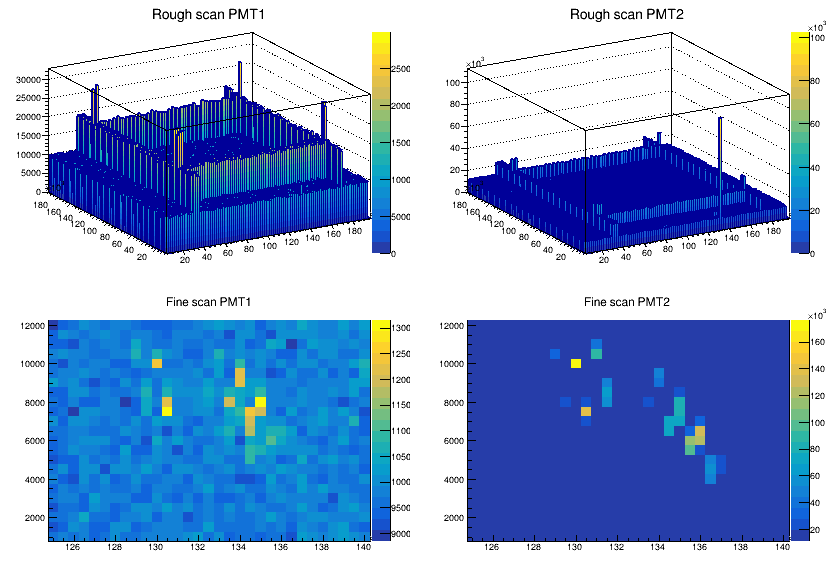}&
    &
    \includegraphics[width=0.48\textwidth]{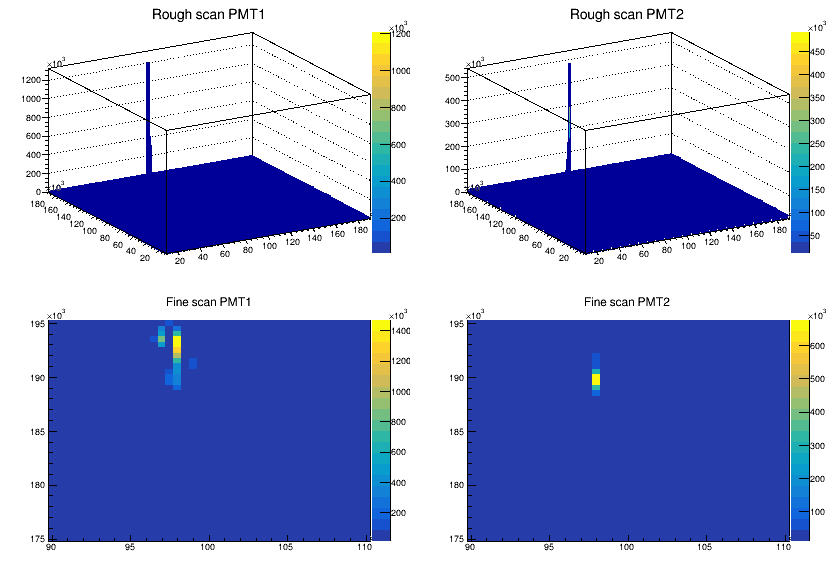}\\
    & & \\
    \includegraphics[width=0.48\textwidth]{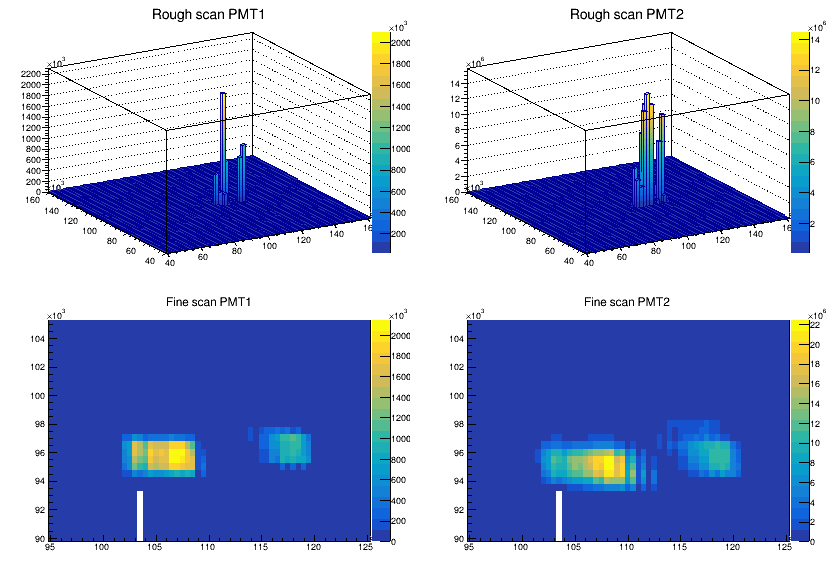}&
    &
    \includegraphics[width=0.48\textwidth]{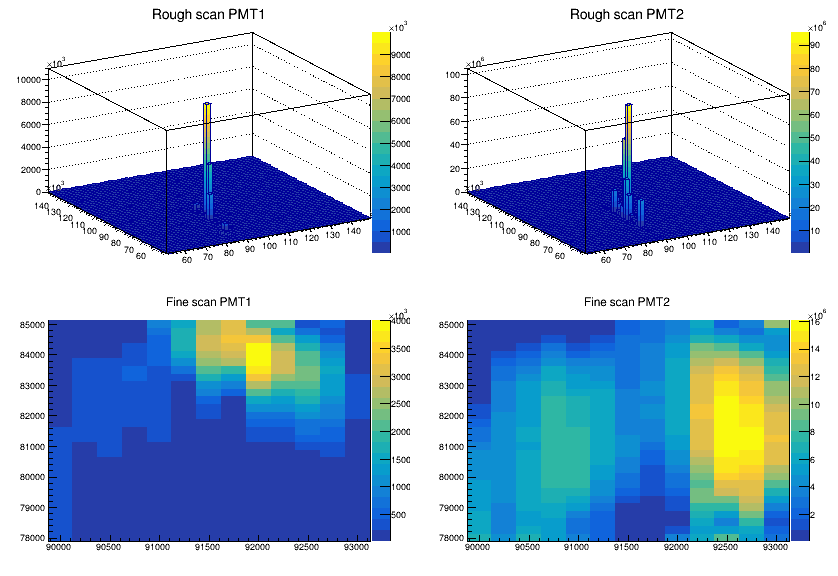}
  \end{tabular}
  \caption{The results of the scans for x- (PMT1) and y- (PMT2)
    polarizations for Oho down \textit{(upper left)}, Oho up
    \textit{(upper right)}, Nikko up \textit{(lower left)}, Nikko down
    \textit{(lower right)}.}
  \label{p:scan-results}
\end{figure}

\begin{figure}[htp]
  \begin{tabular}{cc}
    \includegraphics[width=0.48\textwidth]{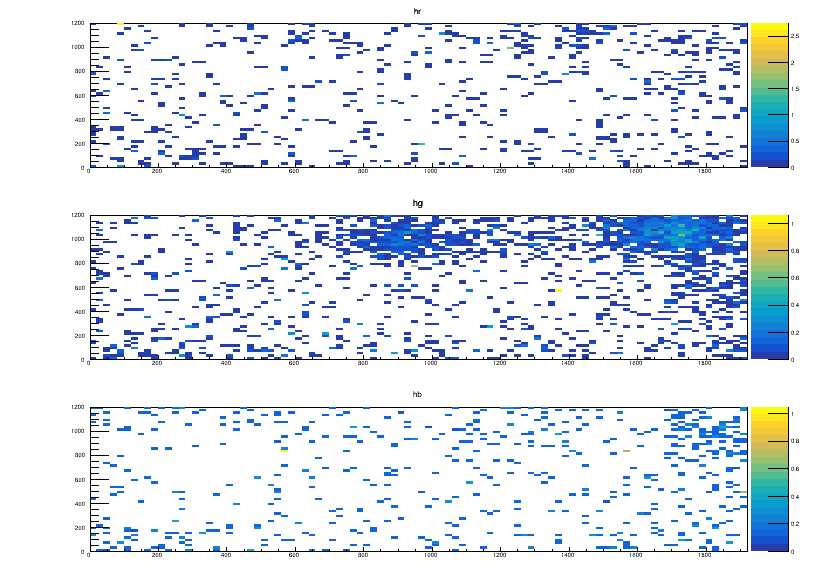}&
    \includegraphics[width=0.48\textwidth]{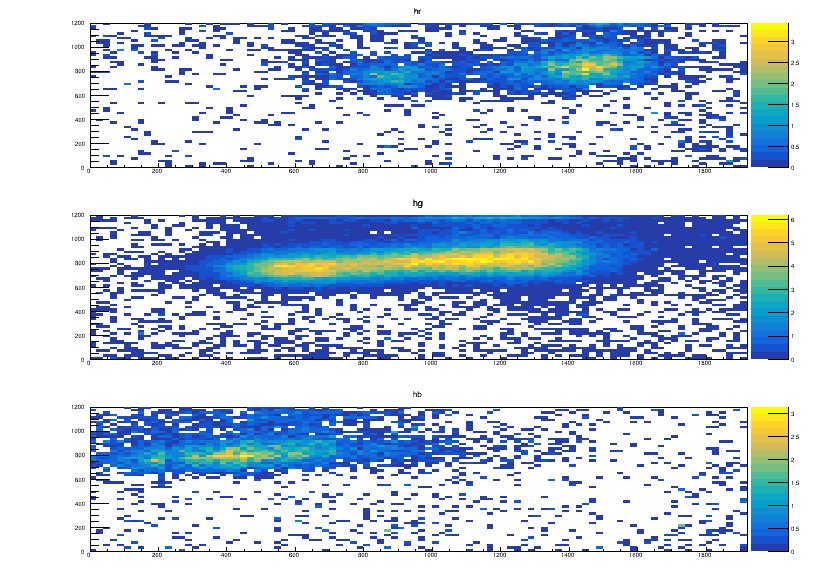}\\
    & \\
    \includegraphics[width=0.48\textwidth]{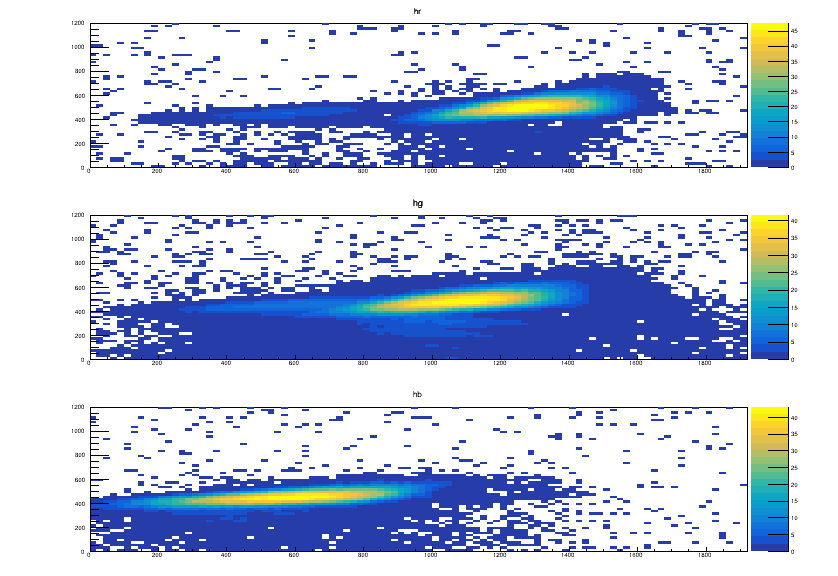}&
    \includegraphics[width=0.48\textwidth]{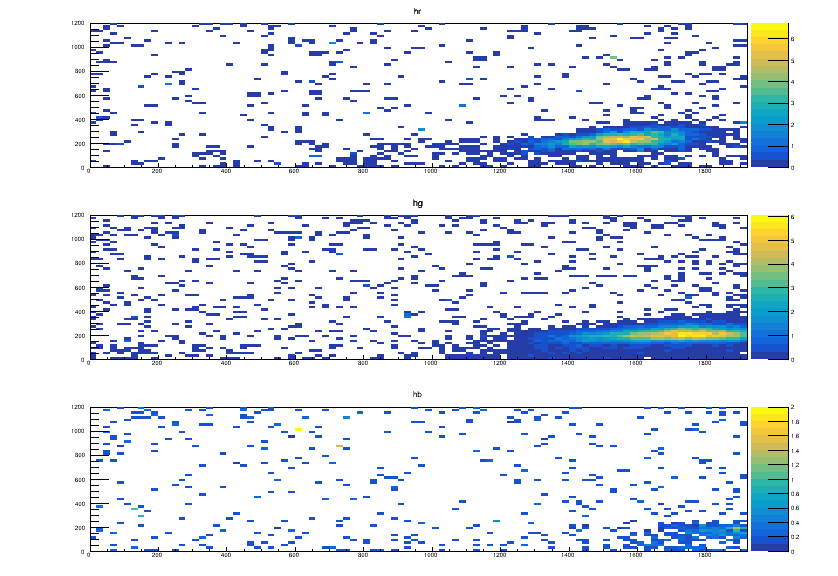}
  \end{tabular}
  \put(-485, 150){R}
  \put(-485,  95){G}
  \put(-485,  40){B}
  \put(-485, -35){R}
  \put(-485, -90){G}
  \put(-485,-145){B}
  \put(-235, 150){R}
  \put(-235,  95){G}
  \put(-235,  40){B}
  \put(-235, -35){R}
  \put(-235, -90){G}
  \put(-235,-145){B}
  \caption{Examples of images of the spot for three color components (R,
    G and B) for Oho down x-polarization \textit{(upper left)}, Oho down
    y-polarization \textit{(upper right)}, Oho up x-polarization
    \textit{(lower left)}, Oho up y-polarization \textit{(lower
      right)}.}
  \label{p:spots-oho}
\end{figure}

\begin{figure}[htp]
  \begin{tabular}{cc}
    \includegraphics[width=0.48\textwidth]{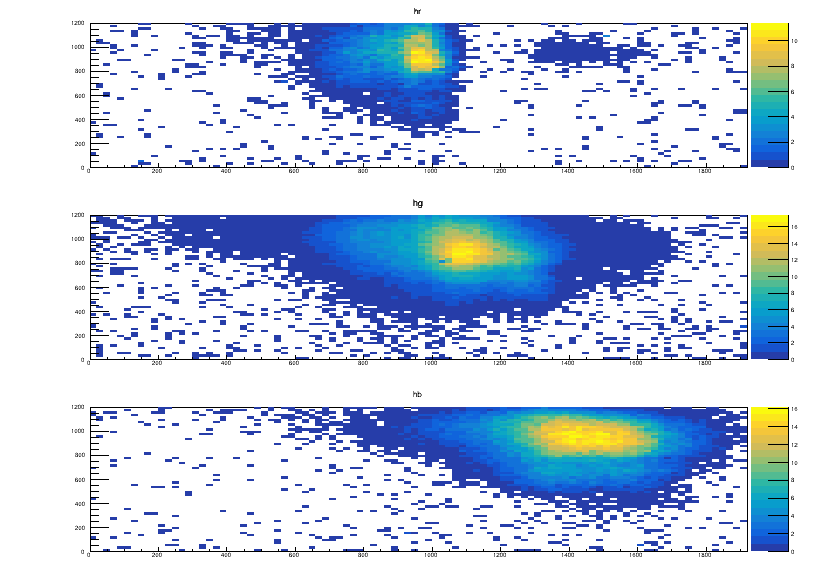}&
    \includegraphics[width=0.48\textwidth]{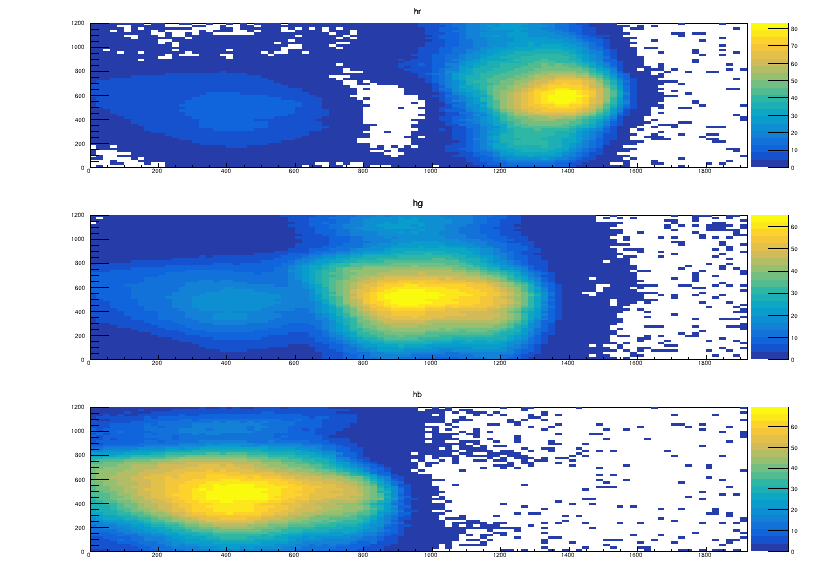}\\
    & \\
    \includegraphics[width=0.48\textwidth]{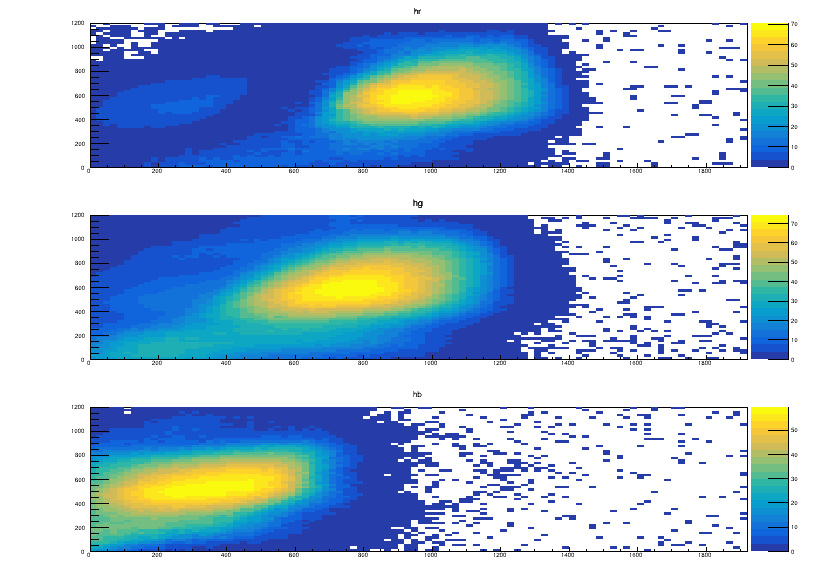}&
    \includegraphics[width=0.48\textwidth]{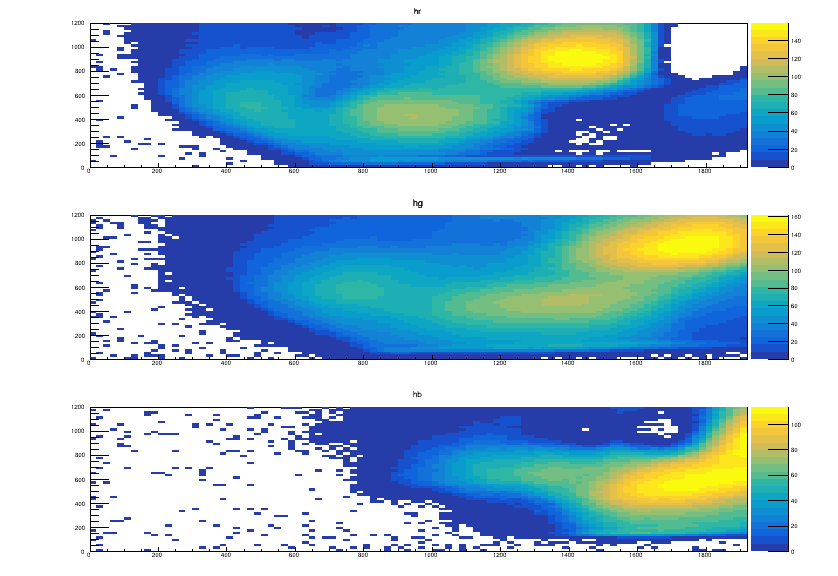}
  \end{tabular}
  \put(-485, 150){R}
  \put(-485,  95){G}
  \put(-485,  40){B}
  \put(-485, -35){R}
  \put(-485, -90){G}
  \put(-485,-145){B}
  \put(-235, 150){R}
  \put(-235,  95){G}
  \put(-235,  40){B}
  \put(-235, -35){R}
  \put(-235, -90){G}
  \put(-235,-145){B}
  \caption{Examples of images of the spot for three color components (R,
    G and B) for Nikko down x-polarization \textit{(upper left)}, Nikko
    down y-polarization \textit{(upper right)}, Nikko up x-polarization
    \textit{(lower left)}, Nikko up y-polarization \textit{(lower
      right)}.}
  \label{p:spots-nikko}
\end{figure}

Before taking data we performed angle scans in search for the light from
the IP, which we call a ``spot''. The results of the scans are shown in
Fig.~\ref{p:scan-results}. For every position of the stepper motors
($x-$ and $y-$axes on the plots) a sum of all pixels of an image was
calculated. This value is shown for both cameras of each telescope as
PMT1 and PMT2. After a rough scan in a wide angle range we performed a
fine scan around a found spot. Due to the hysteresis effect in the
primary mirrors' drives the exact position of the spot may change a
little from a scan to scan, thus final positioning was performed by hand
to fit the spot into a camera sensor. Examples of the spot images for
all eight cameras are shown in Figs.~\ref{p:spots-oho}
and~\ref{p:spots-nikko}. Each image is a combination of three colors:
red, green and blue. As can be seen the position of the spot is slightly
different for different colors. This happens since Wollaston prism is
chromatic and the separation angle between the ordinary and
extraordinary rays depend on the wave length. When a primary mirror
moves the location of the spot changes for two cameras capturing two
light polarizations simultaneously. This and the Wollaston prism
chromaticity limits our ability for the ideal spot positioning.

Data used in this study was taken in May -- June 2024. In total seven
good runs were taken with the total duration of 23 days, one measurement
every eight seconds. All data with a current of at least one beam above
100~mA were considered for analysis.

\section{ML model description}

To obtain the results we again used an ML approach. There were two
methods to process collected images. The first one featured a
convolutional neural network (CNN) which used scaled to $240 \times 192$
pixels but otherwise unprocessed images as input, denoted as ``full
images'' (FI). The second method featured a separate library
\texttt{openCV}~\cite{opencv}, which was used to describe each image
with an array of 768 parameters, such as a position of the brightest
pixel in each spectrum range, its brightness, shapes of the contour
lines, average brightness of the areas outside the spot etc. These
arrays are denoted as ``reduced images'' (RI), and they are used as
input to a NN instead of the images themselves. For most variables the
FI method gives better results than the RI, but it requires much more
computational resources and time for training.

We tried many different models for the FI method. We still continue the
process of refining the model. At the moment of writing the default
model consists of six convolutional layers, combined in pairs followed followed by
an average pooling layer, one fully connected layer and an output
layer. The total number of neurons in this model is approximately
17M. The RI method uses a model with a single hidden fully connected
layer. The total number of neurons in this model is approximately
612K. The models were implemented using Keras~\cite{keras}, a high level
abstraction library that works on top of the low level
TensorFlow~\cite{tensorflow} compute engine.

The data set used consists of about 111000 LABM and SuperKEKB
measurements, which are split into three subsets: about 80000 points
(70\%) for training, 23000 (20\%) for validation, and 11000 (10\%) for
testing of the model. These measurements are randomly shuffled before
being used, meaning that there is no time correlation between successive
points in the data set.

\section{Results}

In this section we show the results obtained with the NN models and 
compare these results with those provided by the traditional Linear
Regression (LR):
\begin{equation}
  y = \beta_0 + \sum_{i} \beta_i x_i\;,
\end{equation}
where independent variables $x_i$ correspond to the number of pixels of
scaled images in FI method and to the number of parameters in an image
description in RI method. Previous study showed feasibility of the LR
approach since the LABM has always seen high linearity with beam current
and other beam parameters in zero beamstrahlung conditions (that is,
when only one beam was present in the accelerator). We also compare our
results with those obtained by the previous LABM configuration
from~\cite{nnresult}.



\subsection{Specific luminosity}

The first variable we try to obtain from our measurements is
luminosity. It is not exactly a beam parameter but it is closely related
to the beam parameters. Also luminosity is one of the two figures of
merit of a collider along with its energy and is the goal of our utmost
interest.

\begin{figure}[htp]
  \begin{tabular}{ccc}
    \includegraphics[width=0.328\textwidth]{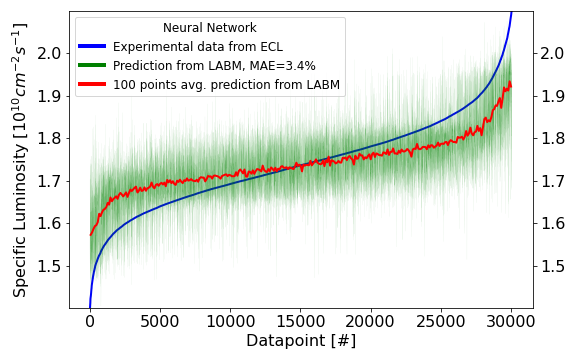}&
    \includegraphics[width=0.3\textwidth]{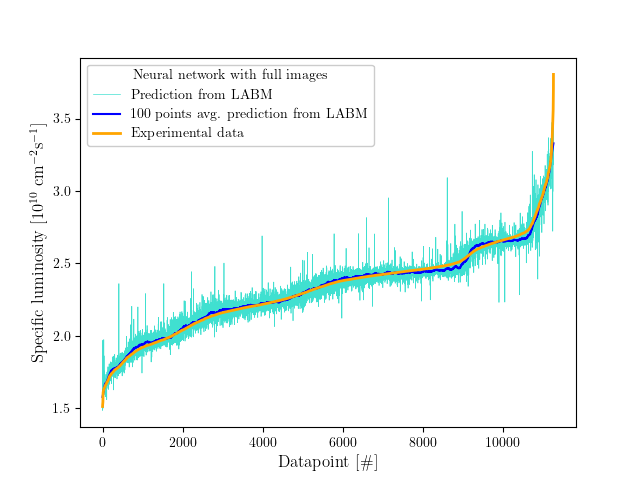}&
    \includegraphics[width=0.3\textwidth]{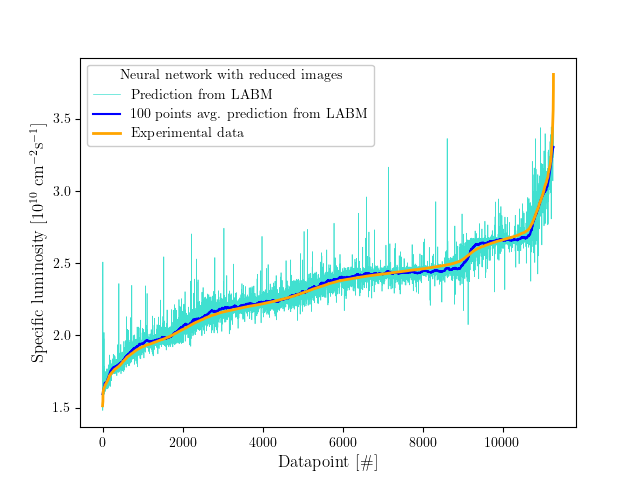}\\
    \includegraphics[width=0.328\textwidth]{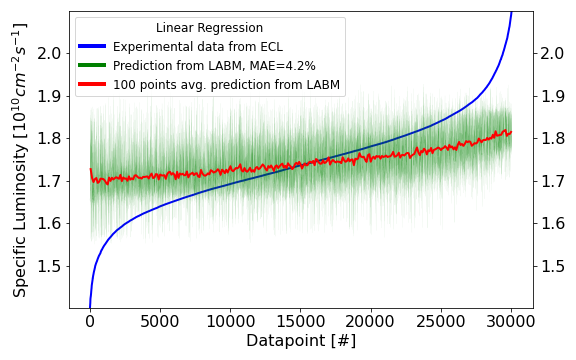}&
    \includegraphics[width=0.3\textwidth]{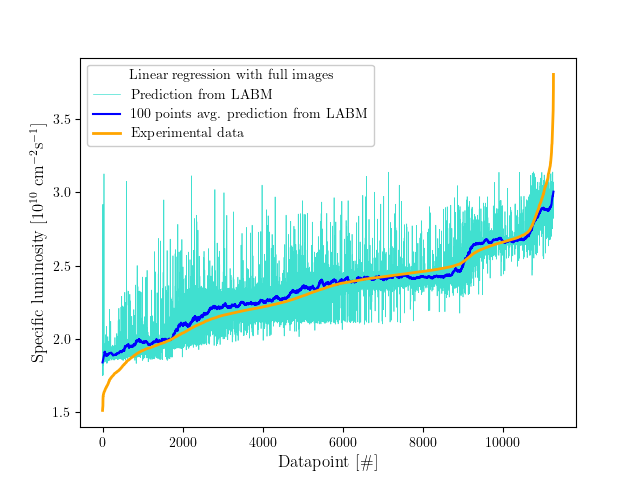}&
    \includegraphics[width=0.3\textwidth]{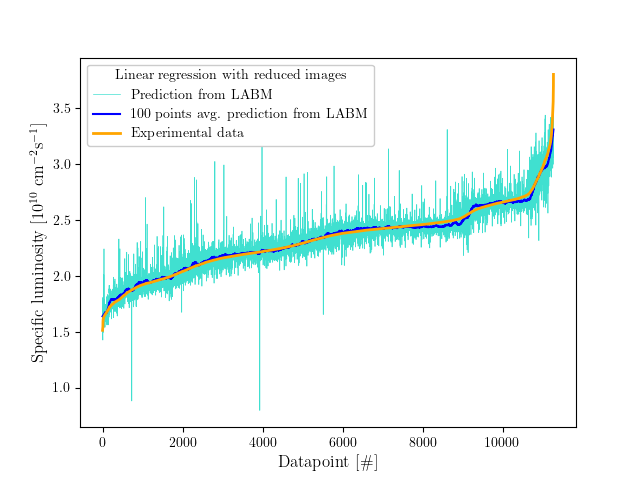}\\
  \end{tabular}
  \caption{Specific luminosity $\mathcal{L}_{spec}$ obtained with the previous
    setup~\cite{nnresult} \textit{(left column)}, and with the upgraded
    setup with full images \textit{(middle column)} and reduced images
    \textit{(right column)} using neural networks \textit{(upper row)}
    and linear regression \textit{(lower row)}. Datapoints are sorted in
    the variable ascending order.}
  \label{p:res-lumspec}
\end{figure}

The absolute luminosity at SuperKEKB is measured by the Electromagnetic
Calorimeter~(ECL) monitor~\cite{ecl}, located in the Belle~II
detector. The ECL measures Bhabha events and, following calibration,
provides an absolute value for the luminosity. For the direct comparison
with the previous configuration we will use the specific luminosity,
defined as:
\begin{equation}
  \mathcal{L}_{s\!p} = \frac{f_0}{2\pi\Sigma_x\Sigma_y}
\end{equation}
where $f_0$ ($0.1\,\mathrm{MHz}$ for SuperKEKB) is the single bunch
revolution frequency and $\Sigma_i$ $(i = x, y)$ are the convoluted beam
sizes, defined as the sum in quadrature of the two beam sizes at the IP:
$\Sigma_i^2 = \sigma_{i,1}^2 + \sigma_{i,2}^2$. It is obtained from the
regular formula of the luminosity by dividing it by the factor $N_b N_1
N_2$, \textit{i.e.} the number of bunches times the product of the
numbers of particles per bunch of the two beams. Defined this way, the
specific luminosity is independent from the beam currents and from the
number of bunches present in the rings. Fig.~\ref{p:res-lumspec} shows
the comparison of the specific luminosity predictions obtained with the
previous and upgraded setups using NN and LR.

\begin{figure}[htp]
  \begin{tabular}{ccc}
    \includegraphics[width=0.328\textwidth]{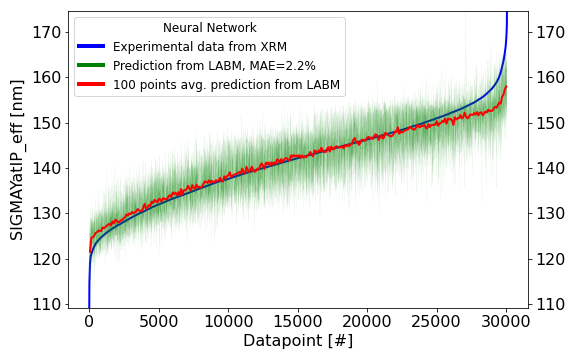}&
    \includegraphics[width=0.3\textwidth]{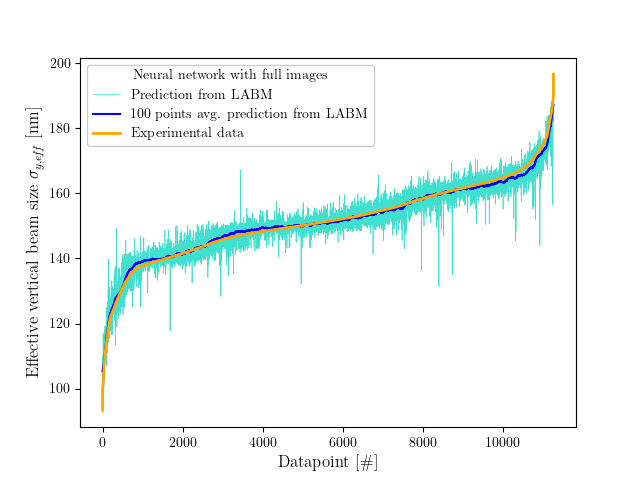}&
    \includegraphics[width=0.3\textwidth]{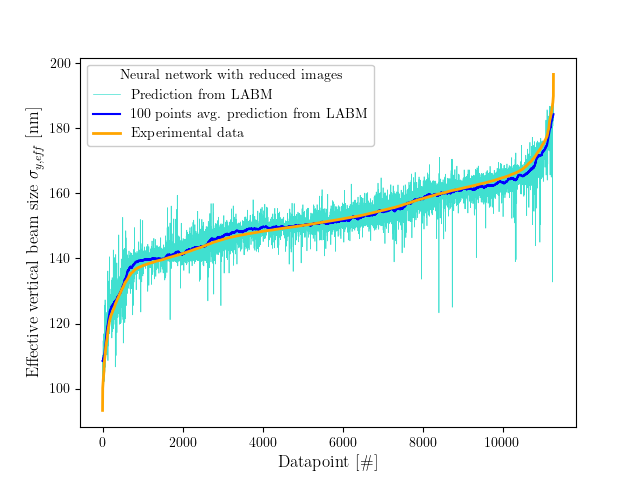}\\
    \includegraphics[width=0.328\textwidth]{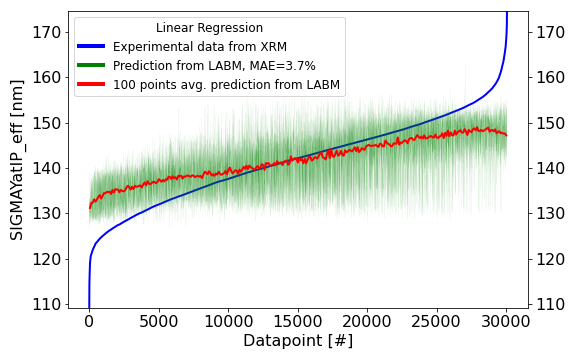}&
    \includegraphics[width=0.3\textwidth]{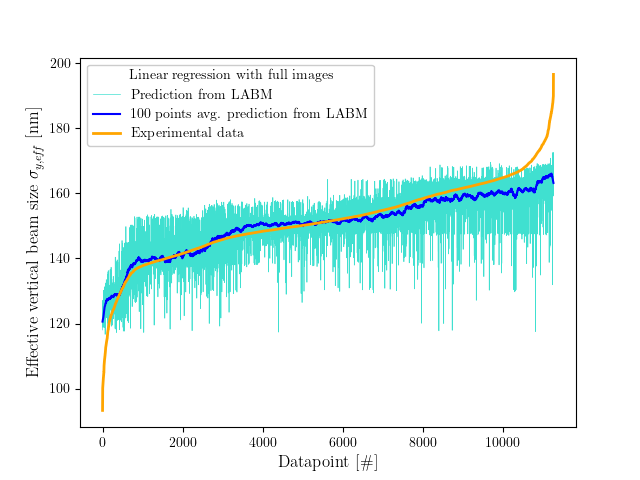}&
    \includegraphics[width=0.3\textwidth]{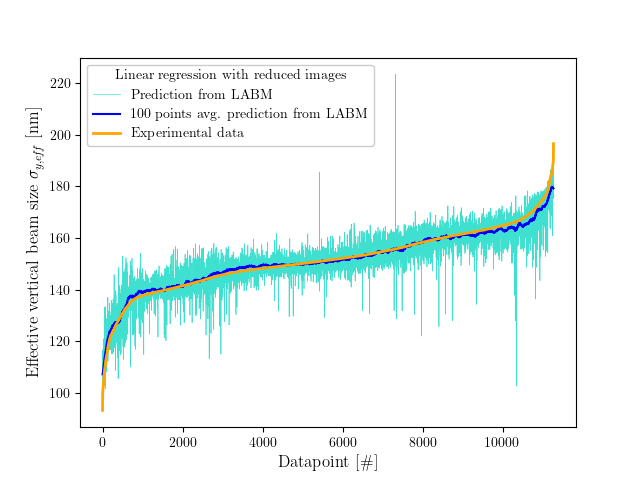}\\
  \end{tabular}
  \caption{Effective vertical beam size $\sigma_{y,e\!f\!f}$ obtained
    with the previous setup~\cite{nnresult} \textit{(left column)}, and
    with the upgraded setup with full images \textit{(middle column)}
    and reduced images \textit{(right column)} using neural networks
    \textit{(upper row)} and linear regression \textit{(lower
      row)}. Datapoints are sorted in the variable ascending order.}
  \label{p:res-sigyeff}
\end{figure}

In the previous publications it was shown that in the SuperKEKB beam
crossing situation the specific luminosity should closely track the
variable
\begin{equation}
  \sigma_{y,e\!f\!f}=\frac{\sigma_{y,1}\sigma_{y,2}}{\Sigma_y}.
\end{equation}
The comparison of the results for this variable obtained with the
previous and upgraded setups is shown in Fig.~\ref{p:res-sigyeff}.

\subsection{Beams sizes}

The beams sizes at SuperKEKB is measured by the X-ray monitor
(XRM)~\cite{xrm}. Two XRM monitors are installed at the HER ($e^-$) and
LER ($e^+$) rings, at 641.4$\,$m and 1397.7$\,$m from the IP,
respectively. Using the Twiss parameters at their location, it is
possible to estimate of the emittance, and through the optical transfer
matrices it is possible to obtain an estimation of the beams sizes at
the IP, which are the quantities we are interested in.

\begin{figure}[htp]
  \begin{tabular}{ccc}
    \includegraphics[width=0.328\textwidth]{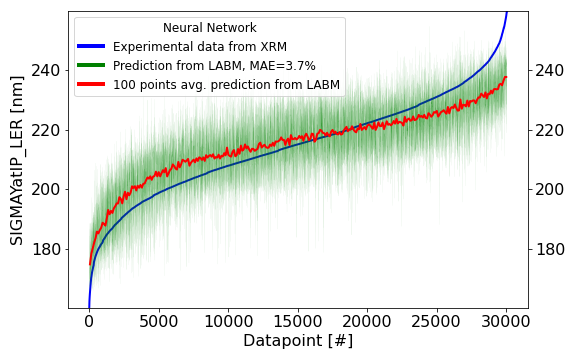}&
    \includegraphics[width=0.3\textwidth]{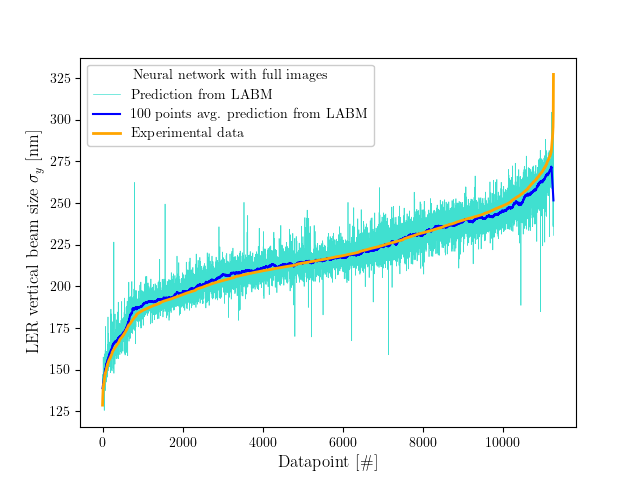}&
    \includegraphics[width=0.3\textwidth]{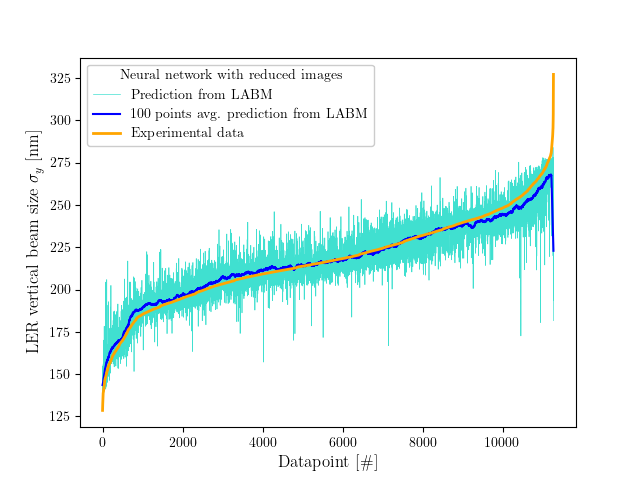}\\
    \includegraphics[width=0.328\textwidth]{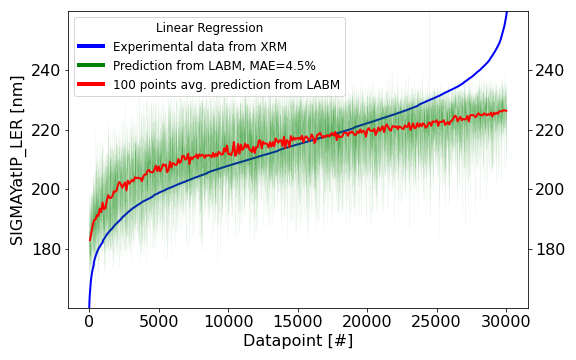}&
    \includegraphics[width=0.3\textwidth]{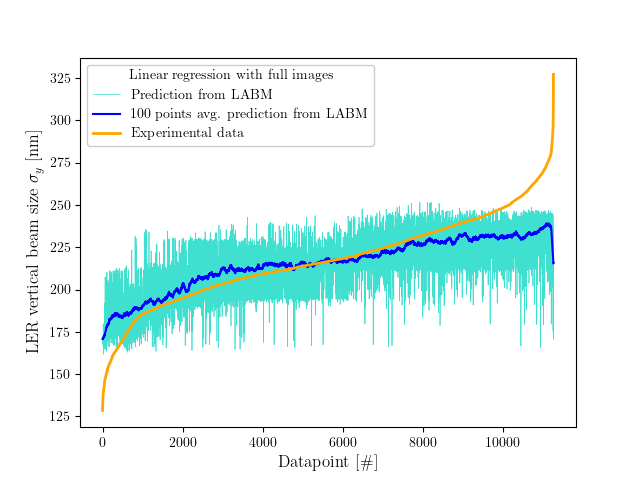}&
    \includegraphics[width=0.3\textwidth]{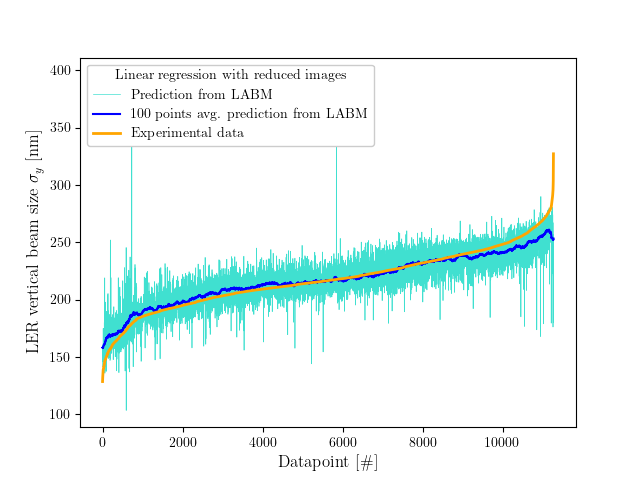}\\
  \end{tabular}
  \caption{LER vertical size $\sigma_{y,LER}$ obtained with the previous
    setup~\cite{nnresult} \textit{(left column)}, and with the upgraded
    setup with full images \textit{(middle column)} and reduced images
    \textit{(right column)} using neural networks \textit{(upper row)}
    and linear regression \textit{(lower row)}. Datapoints are sorted in
    the variable ascending order.}
  \label{p:res-lsigy}
\end{figure}

\begin{figure}[htp]
  \begin{tabular}{ccc}
    \includegraphics[width=0.328\textwidth]{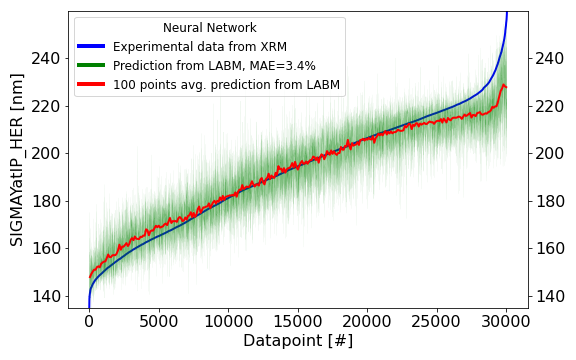}&
    \includegraphics[width=0.3\textwidth]{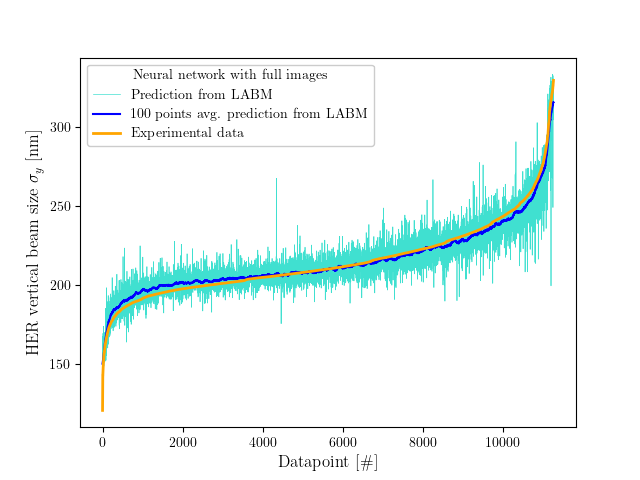}&
    \includegraphics[width=0.3\textwidth]{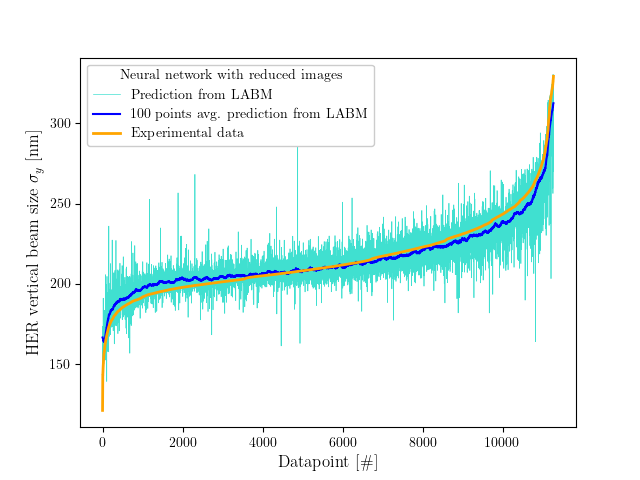}\\
    \includegraphics[width=0.328\textwidth]{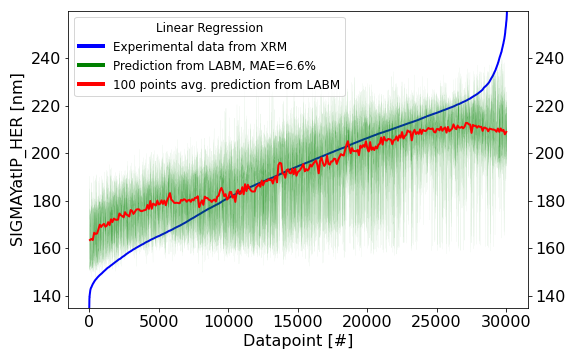}&
    \includegraphics[width=0.3\textwidth]{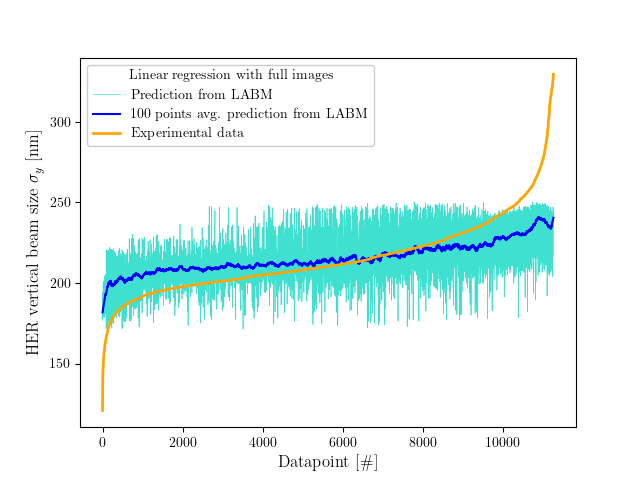}&
    \includegraphics[width=0.3\textwidth]{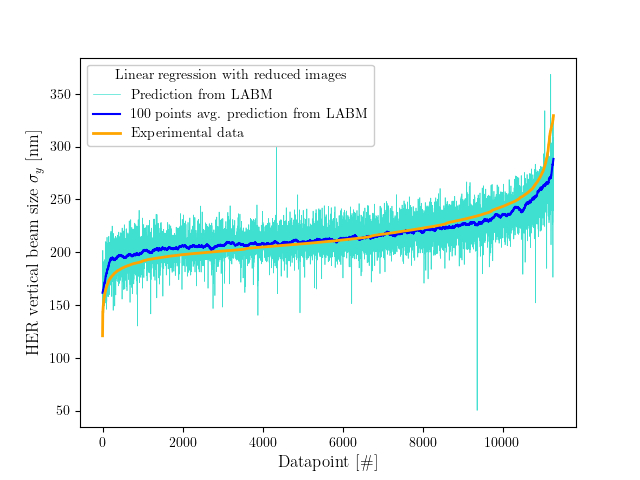}\\
  \end{tabular}
  \caption{HER vertical size $\sigma_{y,HER}$ obtained with the previous
    setup~\cite{nnresult} \textit{(left column)}, and with the upgraded setup
    with full images CNN \textit{(middle column)} and reduced images NN
    \textit{(right column)}. Datapoints are sorted in the variable ascending
    order.}
  \label{p:res-hsigy}
\end{figure}

\begin{figure}[htp]
  \begin{tabular}{ccc}
    \includegraphics[width=0.328\textwidth]{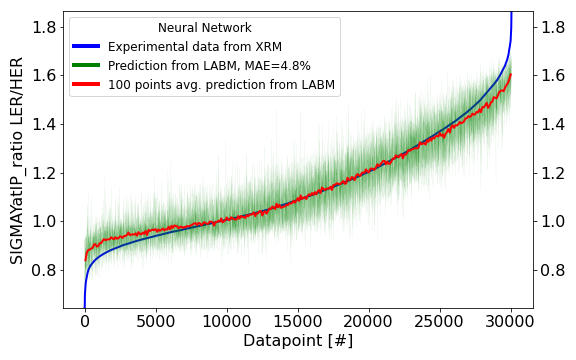}&
    \includegraphics[width=0.3\textwidth]{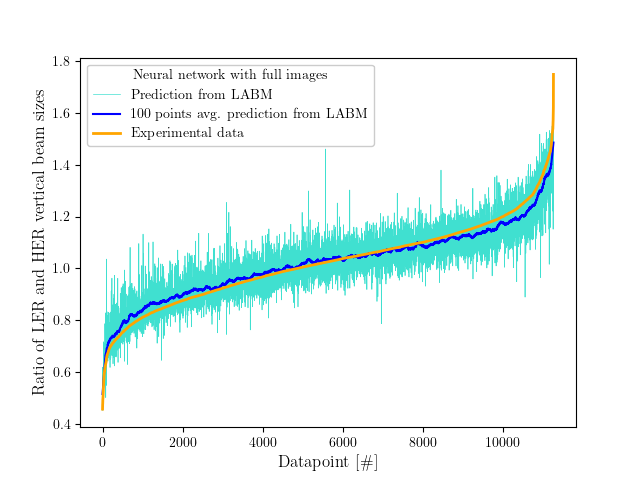}&
    \includegraphics[width=0.3\textwidth]{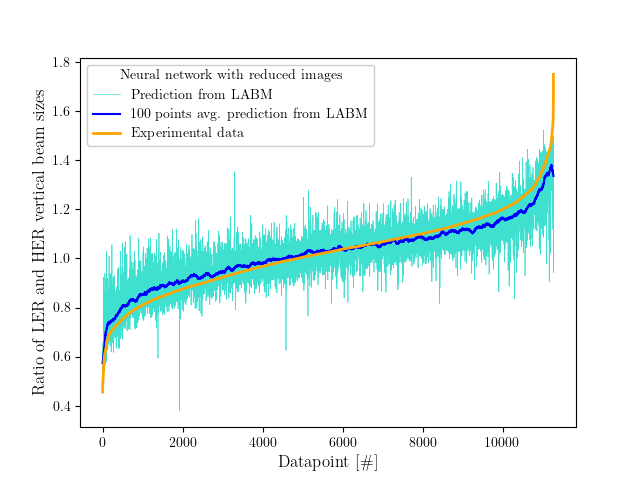}\\
    \includegraphics[width=0.328\textwidth]{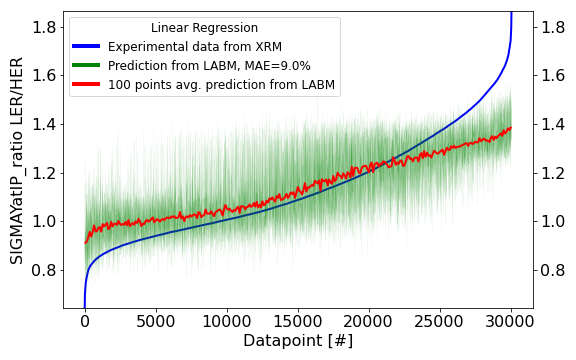}&
    \includegraphics[width=0.3\textwidth]{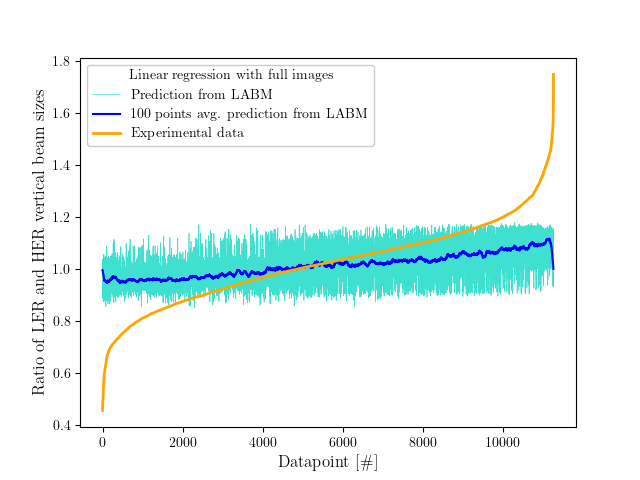}&
    \includegraphics[width=0.3\textwidth]{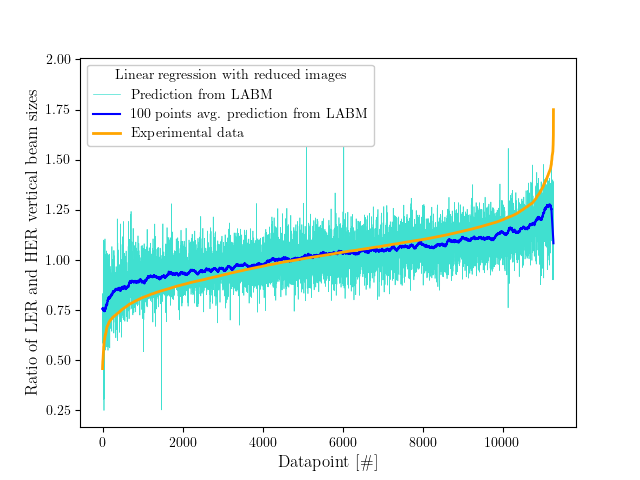}\\
  \end{tabular}
  \caption{Ratio of LER and HER vertical sizes
    $\sigma_{y,LER}/\sigma_{y,HER}$ obtained with the previous
    setup~\cite{nnresult} \textit{(left column)}, and with the upgraded
    setup with full images \textit{(middle column)} and reduced images
    \textit{(right column)} using neural networks \textit{(upper row)}
    and linear regression \textit{(lower row)}. Datapoints are sorted in
    the variable ascending order.}
  \label{p:res-lhsigy}
\end{figure}

\begin{figure}[htp]
  \begin{tabular}{ccc}
    \includegraphics[width=0.328\textwidth]{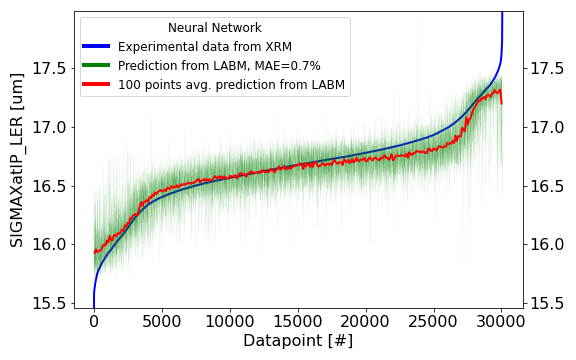}&
    \includegraphics[width=0.3\textwidth]{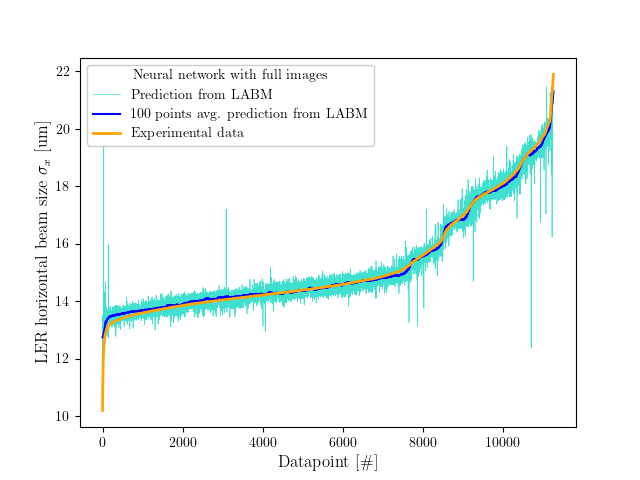}&
    \includegraphics[width=0.3\textwidth]{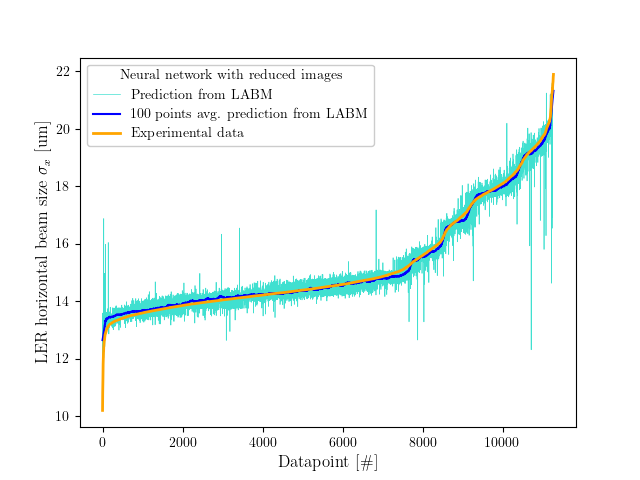}\\
    \includegraphics[width=0.328\textwidth]{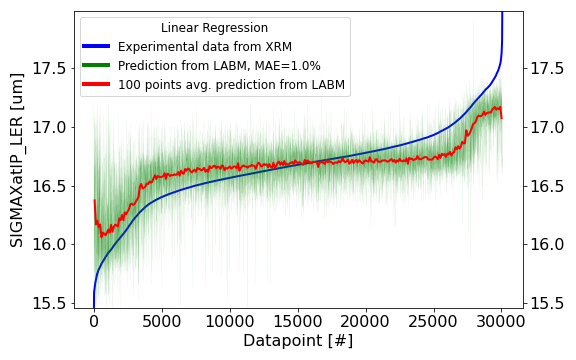}&
    \includegraphics[width=0.3\textwidth]{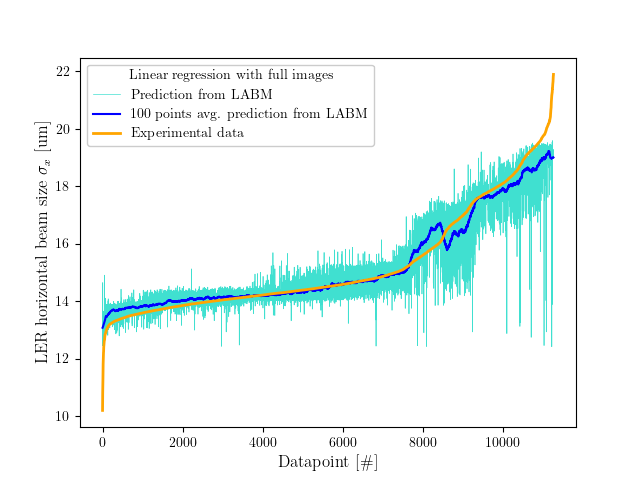}&
    \includegraphics[width=0.3\textwidth]{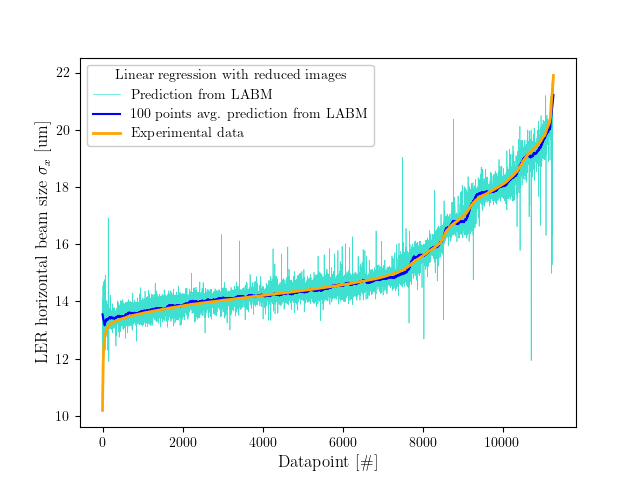}\\
  \end{tabular}
  \caption{LER horizontal size $\sigma_{x,LER}$ obtained with the
    previous setup~\cite{nnresult} \textit{(left column)}, and with the
    upgraded setup with full images \textit{(middle column)} and reduced
    images \textit{(right column)} using neural networks \textit{(upper
      row)} and linear regression \textit{(lower row)}. Datapoints are
    sorted in the variable ascending order.}
  \label{p:res-lsigx}
\end{figure}

\begin{figure}[htp]
  \begin{tabular}{ccc}
    \includegraphics[width=0.328\textwidth]{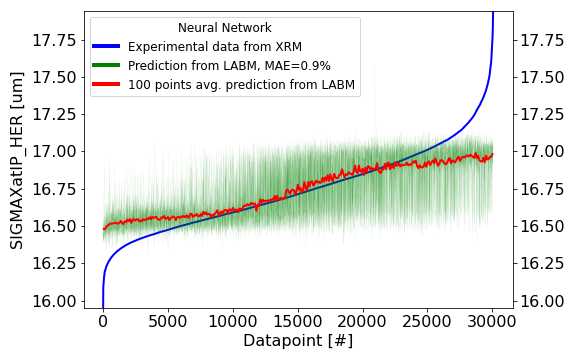}&
    \includegraphics[width=0.3\textwidth]{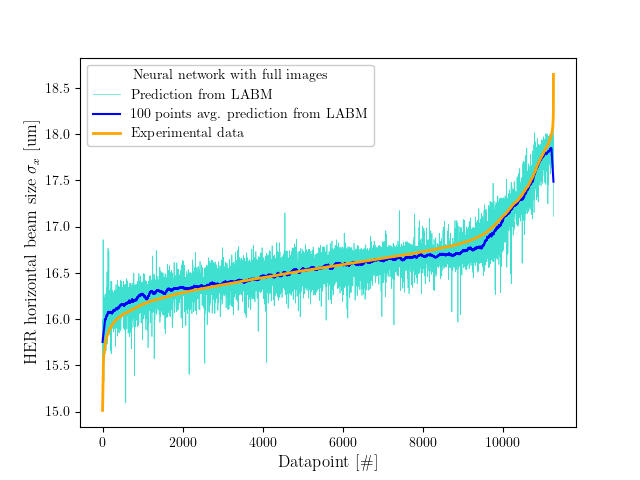}&
    \includegraphics[width=0.3\textwidth]{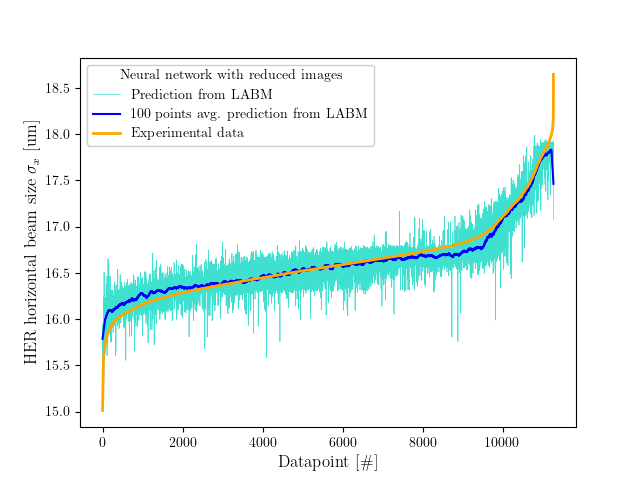}\\
    \includegraphics[width=0.328\textwidth]{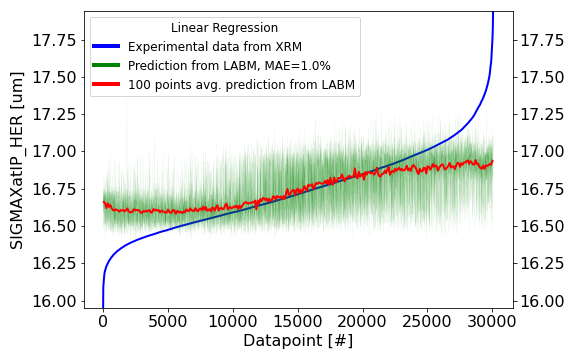}&
    \includegraphics[width=0.3\textwidth]{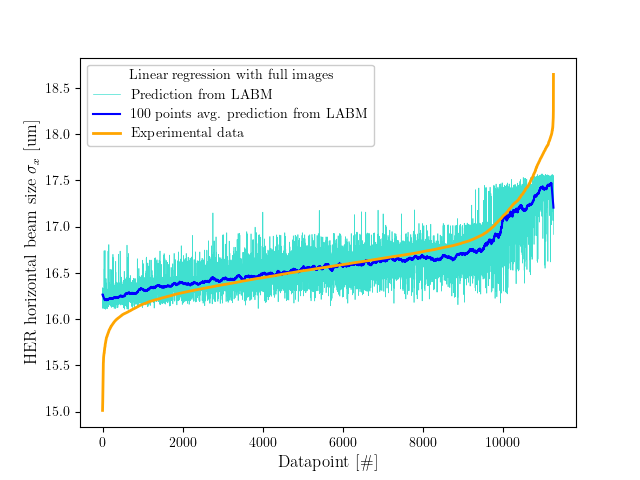}&
    \includegraphics[width=0.3\textwidth]{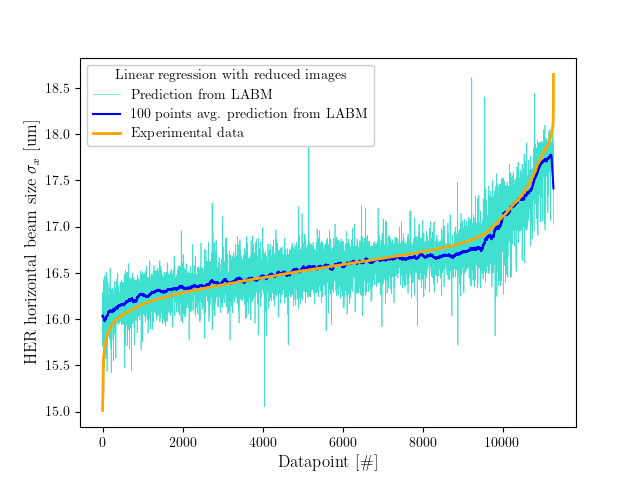}\\
  \end{tabular}
  \caption{HER horizontal size $\sigma_{x,HER}$ obtained with the
    previous setup~\cite{nnresult} \textit{(left column)}, and with the
    upgraded setup with full images \textit{(middle column)} and reduced
    images \textit{(right column)}. Datapoints are sorted in the
    variable ascending order.}
  \label{p:res-hsigx}
\end{figure}

\begin{figure}[htp]
  \begin{tabular}{ccc}
    \includegraphics[width=0.328\textwidth]{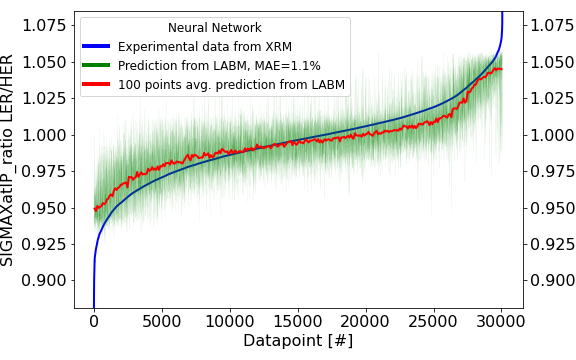}&
    \includegraphics[width=0.3\textwidth]{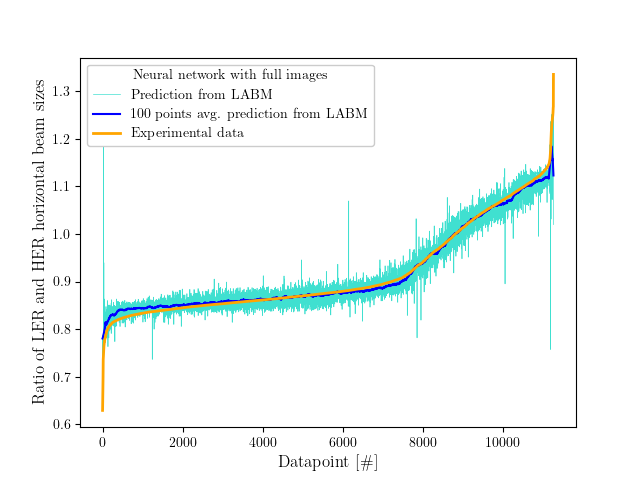}&
    \includegraphics[width=0.3\textwidth]{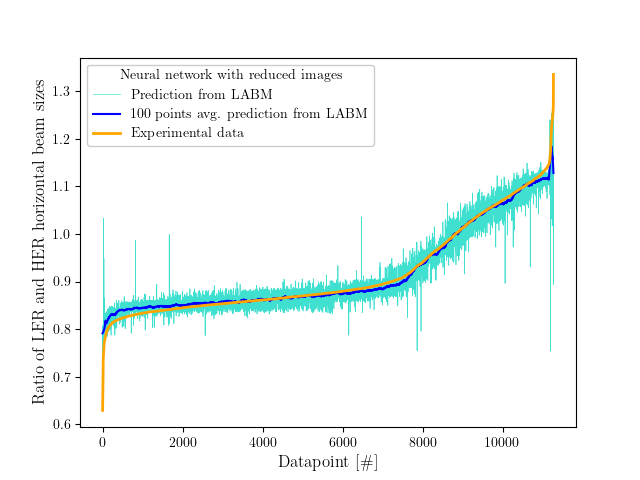}\\
    \includegraphics[width=0.328\textwidth]{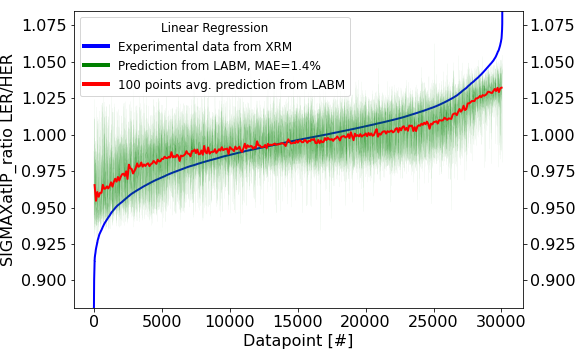}&
    \includegraphics[width=0.3\textwidth]{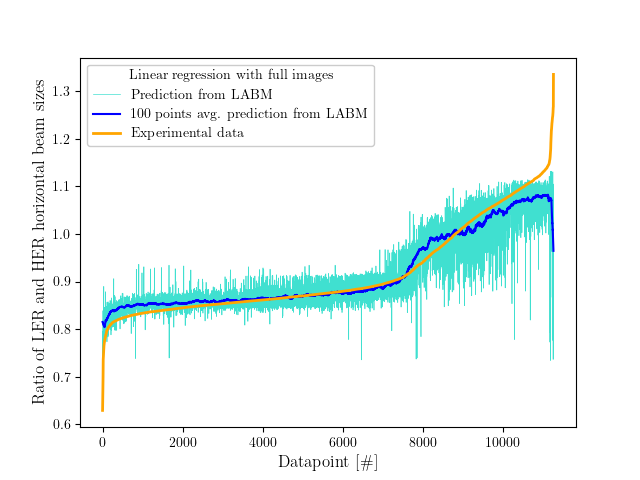}&
    \includegraphics[width=0.3\textwidth]{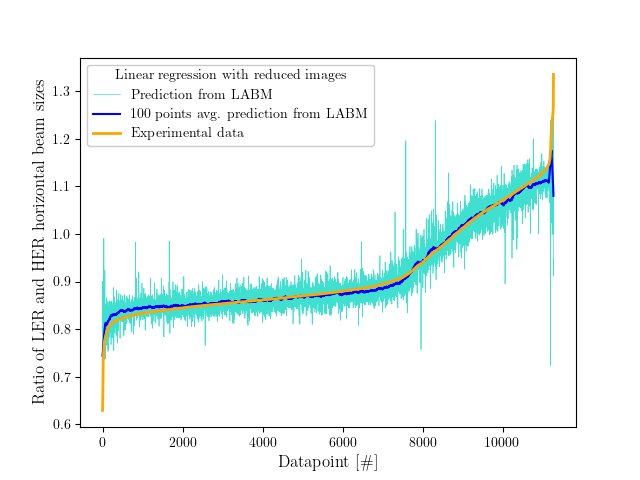}\\
  \end{tabular}
  \caption{Ratio of LER and HER horizontal sizes
    $\sigma_{x,LER}/\sigma_{x,HER}$ obtained with the previous
    setup~\cite{nnresult} \textit{(left column)}, and with the upgraded
    setup with full images \textit{(middle column)} and reduced images
    \textit{(right column)} using neural networks \textit{(upper row)}
    and linear regression \textit{(lower row)}. Datapoints are sorted in
    the variable ascending order.}
  \label{p:res-lhsigx}
\end{figure}

Figs.~\ref{p:res-lsigy} and~\ref{p:res-hsigy} show the results for the
vertical sizes of the beams $\sigma_y$. We are also interested in the
LER/HER ratio on $\sigma_y$ at IP. In fact, KEKB had a vertical beam
size for the LER that was consistently larger than the corresponding one
for the HER. This corresponds to one beam being unfocused, causing
luminosity degradation, which is an effect that we want to prevent at
SuperKEKB. Fig.~\ref{p:res-lhsigy} shows the ratio of the LER and HER
vertical sizes $\sigma_y$. The plots related to horizontal beam widths
and their ratio are shown in Figs.~\ref{p:res-lsigx}
to~\ref{p:res-lhsigx}.

\subsection{Discussion}

The comparison of the obtained accuracy with the previous
publication~\cite{nnresult} is shown in Table~\ref{t:comp}. Generally
the upgraded setup could deliver a better result compared to the
previous setup, though the result for some variables is worse, notably
the LER horizontal size $\sigma_{x,LER}$ and the ratio of horizontal
sizes $\sigma_{x,LER}/\sigma_{x,HER}$. If we compare the results
obtained by two different NNs with the upgraded setup, the FI CNN is
more precise. It is interesting to note that if we compare the LR
results, it is the opposite and the RI LR result is consistently better
than the FI LR. Since obviously the FI and RI methods pick up different
characteristics of the images, it may be beneficial to combine these two
methods into one and use a hybrid NN in the future, which could use the
best from both of them.

\begin{figure}[htp]
  \begin{tabular}{cc}
    \includegraphics[width=0.49\textwidth,trim=1.7cm 0.2cm 1.7cm 0.2cm,clip=true]{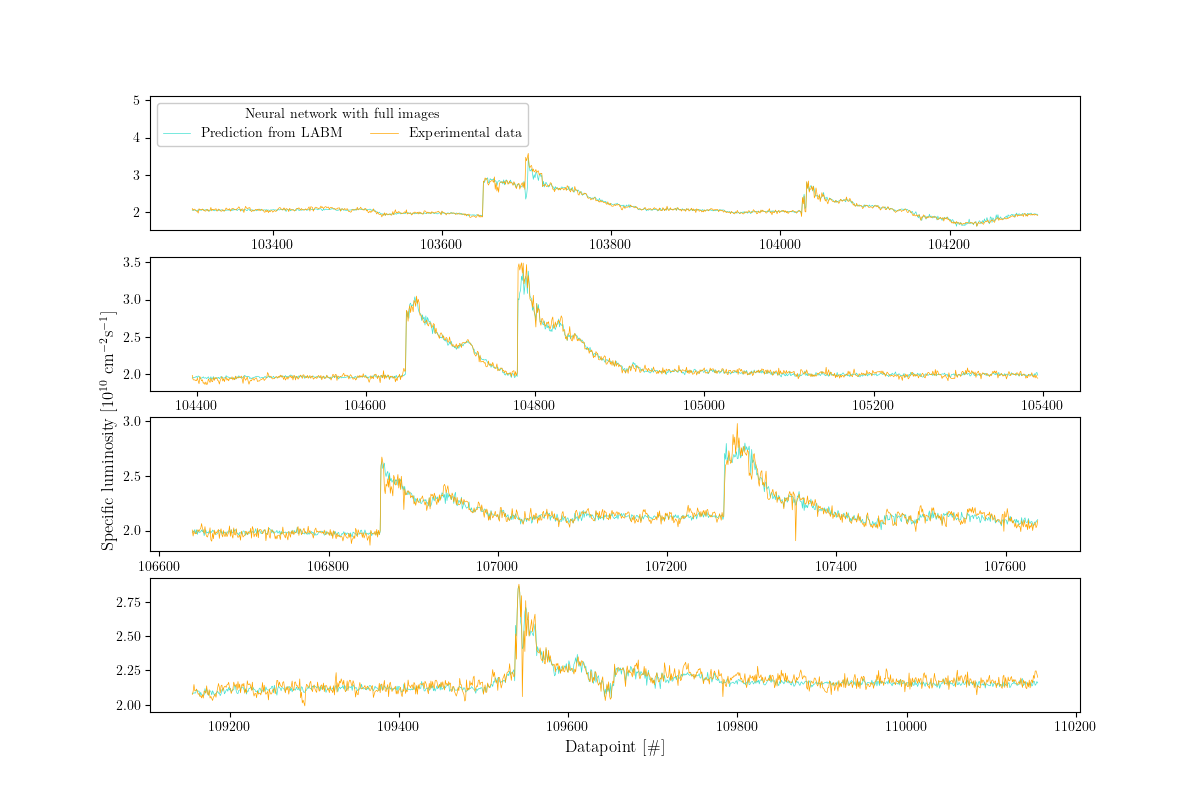}&
    \includegraphics[width=0.49\textwidth,trim=1.7cm 0.2cm 1.7cm 0.2cm,clip=true]{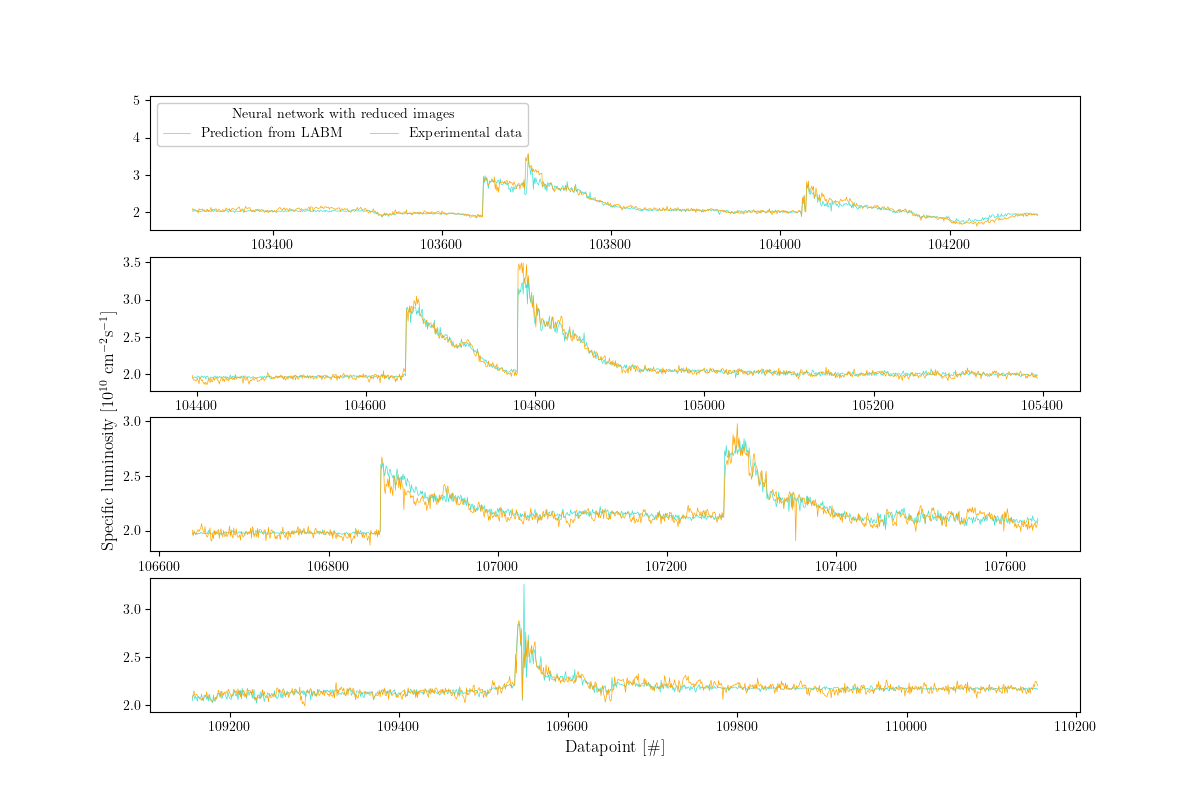}\\
  \end{tabular}
  \caption{Specific luminosity $\mathcal{L}_{spec}$ obtained with the upgraded setup with
    full images CNN \textit{(left)} and reduced images NN
    \textit{(right)}. Datapoints are sorted in chronological order.}
  \label{p:res-lumspec-chrono}
\end{figure}

\begin{figure}[htp]
  \begin{tabular}{cc}
    \includegraphics[width=0.49\textwidth,trim=1.7cm 0.2cm 1.7cm 0.2cm,clip=true]{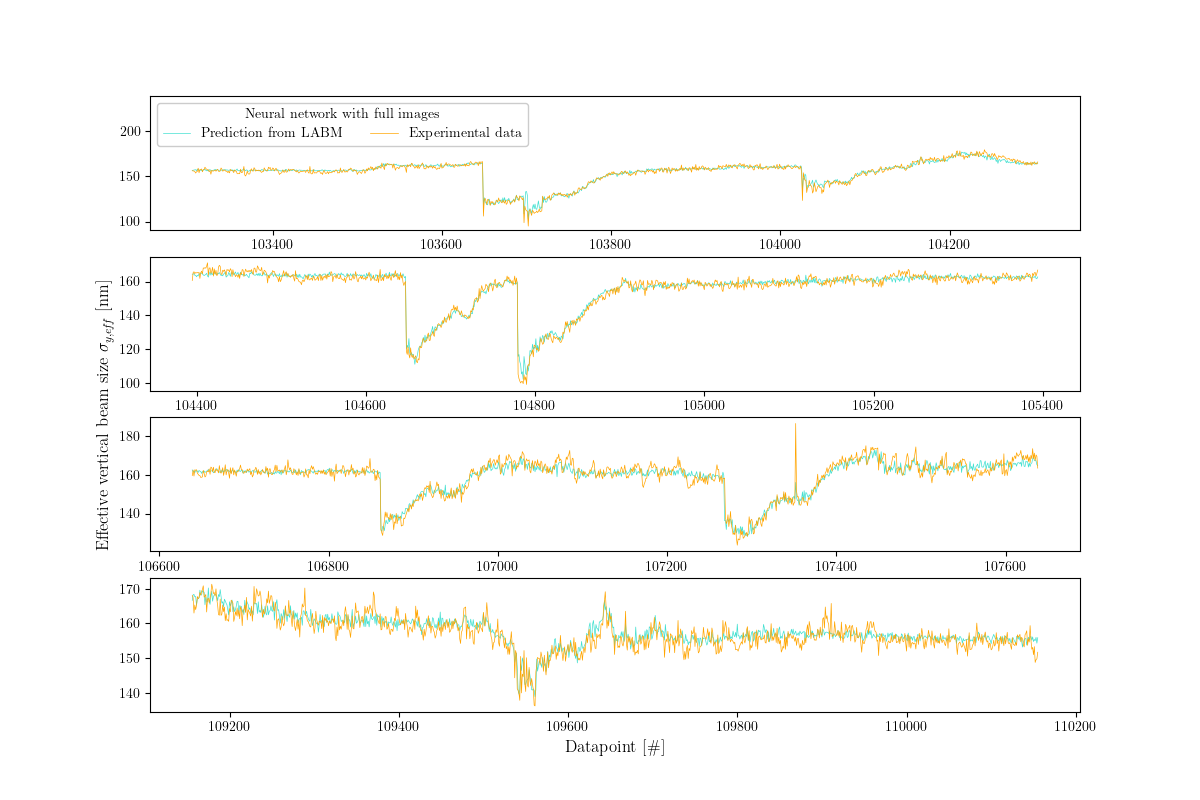}&
    \includegraphics[width=0.49\textwidth,trim=1.7cm 0.2cm 1.7cm 0.2cm,clip=true]{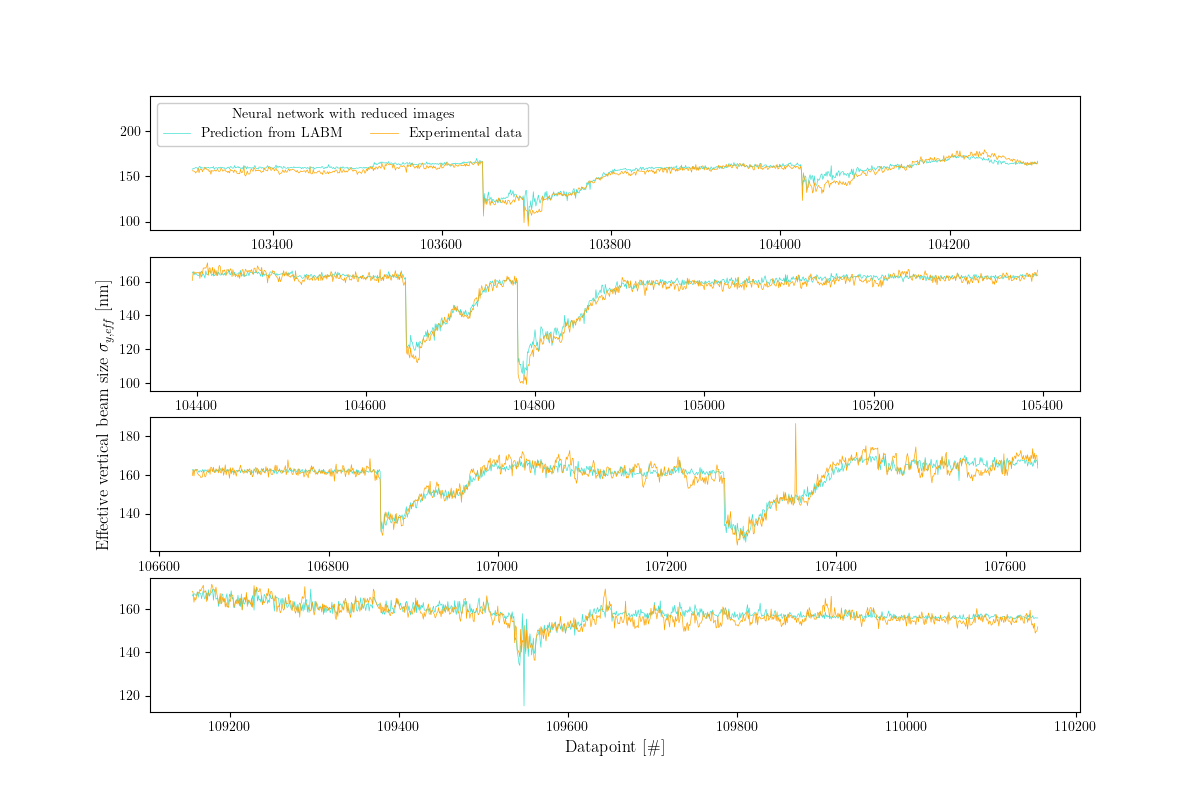}\\
  \end{tabular}
  \caption{Effective vertical beam size $\sigma_{y,e\!f\!f}$ obtained
    with the upgraded setup with full images CNN \textit{(left)} and
    reduced images NN \textit{(right)}. Datapoints are sorted in
    chronological order.}
  \label{p:res-sigyeff-chrono}
\end{figure}

\begin{figure}[htp]
  \begin{tabular}{cc}
    \includegraphics[width=0.49\textwidth,trim=1.7cm 0.2cm 1.7cm 0.2cm,clip=true]{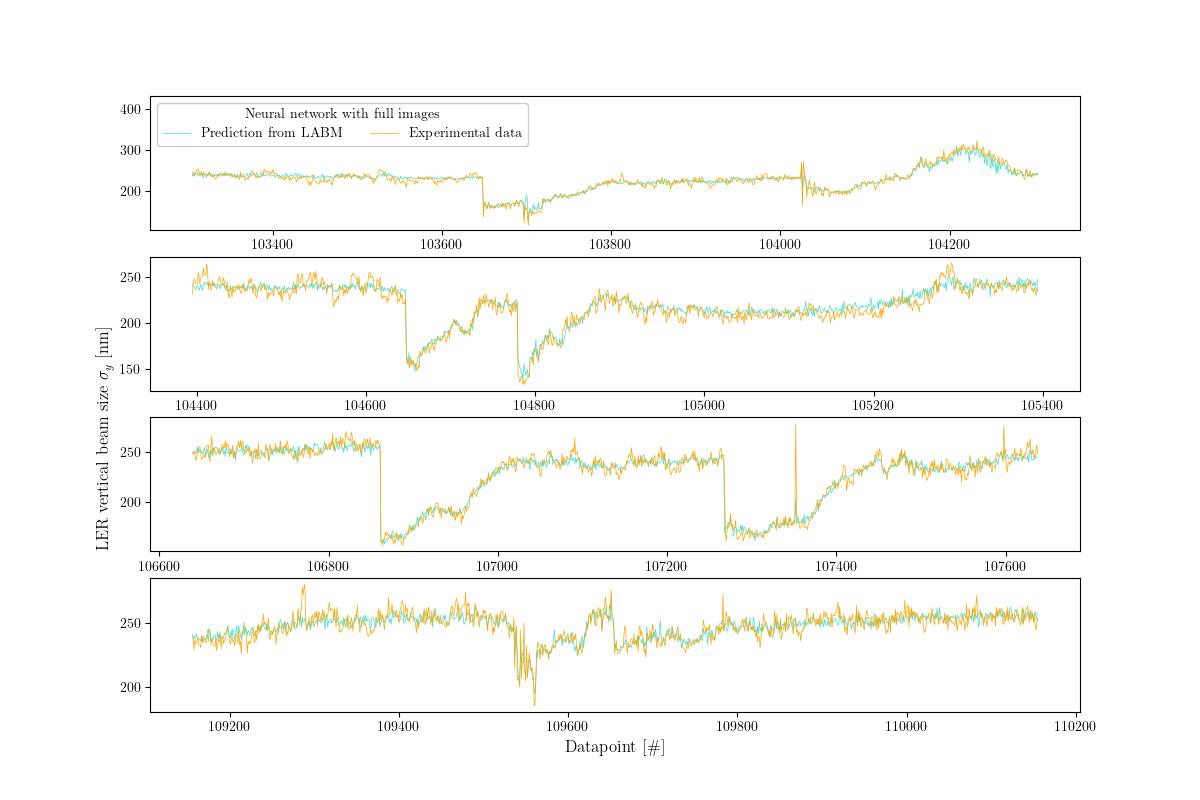}&
    \includegraphics[width=0.49\textwidth,trim=1.7cm 0.2cm 1.7cm 0.2cm,clip=true]{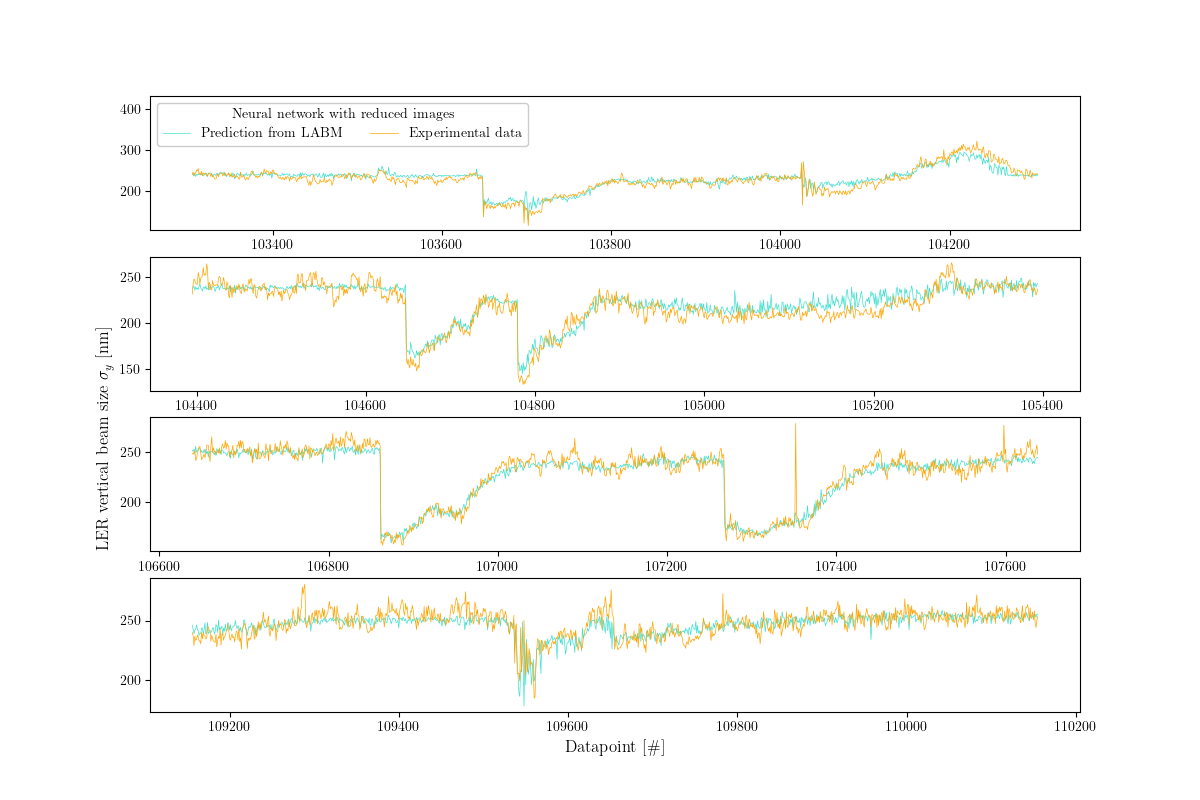}\\
  \end{tabular}
  \caption{LER vertical size $\sigma_{y,LER}$ obtained with the upgraded
    setup with full images CNN \textit{(left)} and reduced images NN
    \textit{(right)}. Datapoints are sorted in chronological order.}
  \label{p:res-lsigy-chrono}
\end{figure}

\begin{figure}[htp]
  \begin{tabular}{cc}
    \includegraphics[width=0.49\textwidth,trim=1.7cm 0.2cm 1.7cm 0.2cm,clip=true]{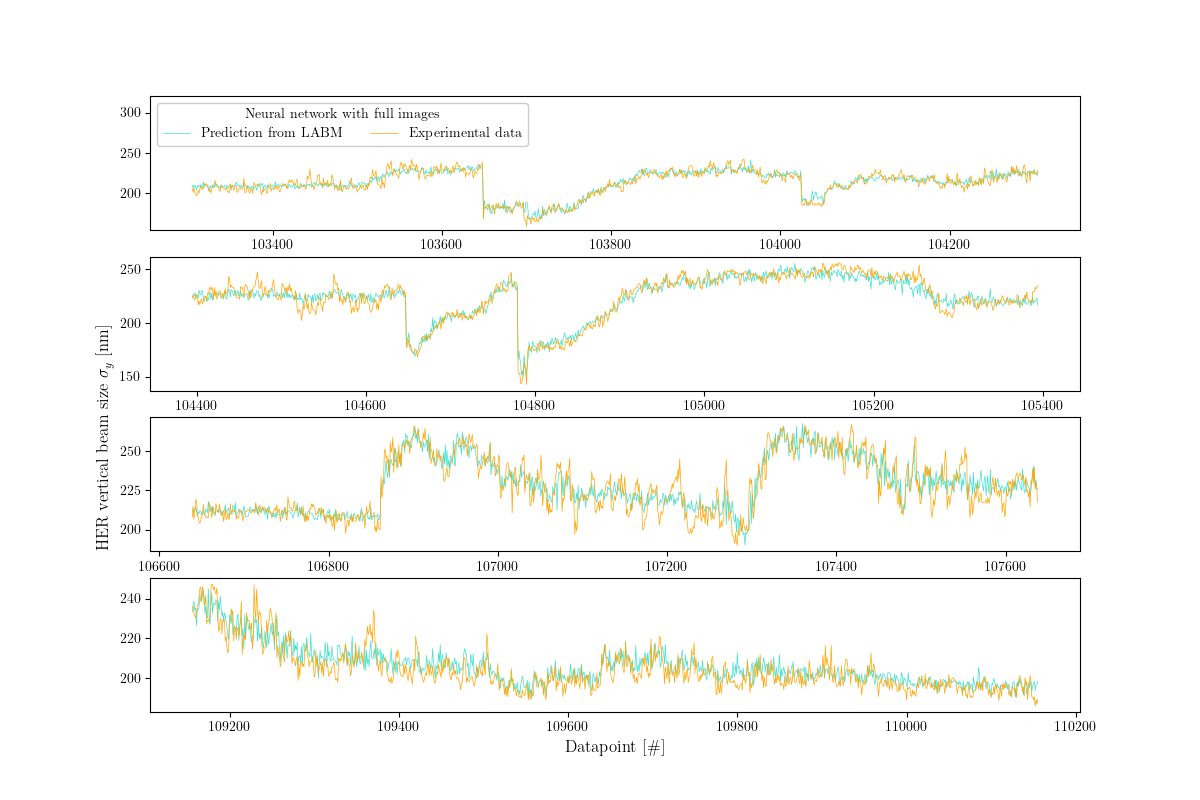}&
    \includegraphics[width=0.49\textwidth,trim=1.7cm 0.2cm 1.7cm 0.2cm,clip=true]{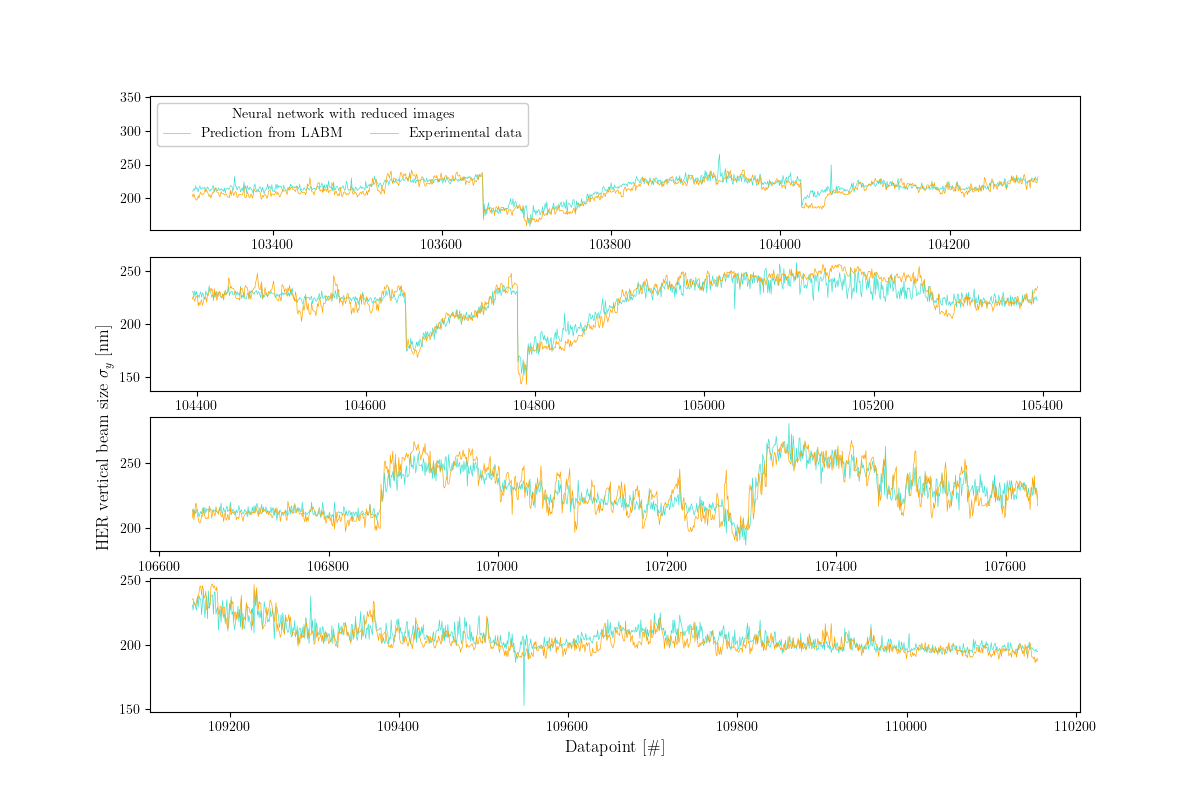}\\
  \end{tabular}
  \caption{HER vertical size $\sigma_{y,HER}$ obtained with the upgraded
    setup with full images CNN \textit{(left)} and reduced images NN
    \textit{(right)}. Datapoints are sorted in chronological order.}
  \label{p:res-hsigy-chrono}
\end{figure}

\begin{figure}[htp]
  \begin{tabular}{cc}
    \includegraphics[width=0.49\textwidth,trim=1.7cm 0.2cm 1.7cm 0.2cm,clip=true]{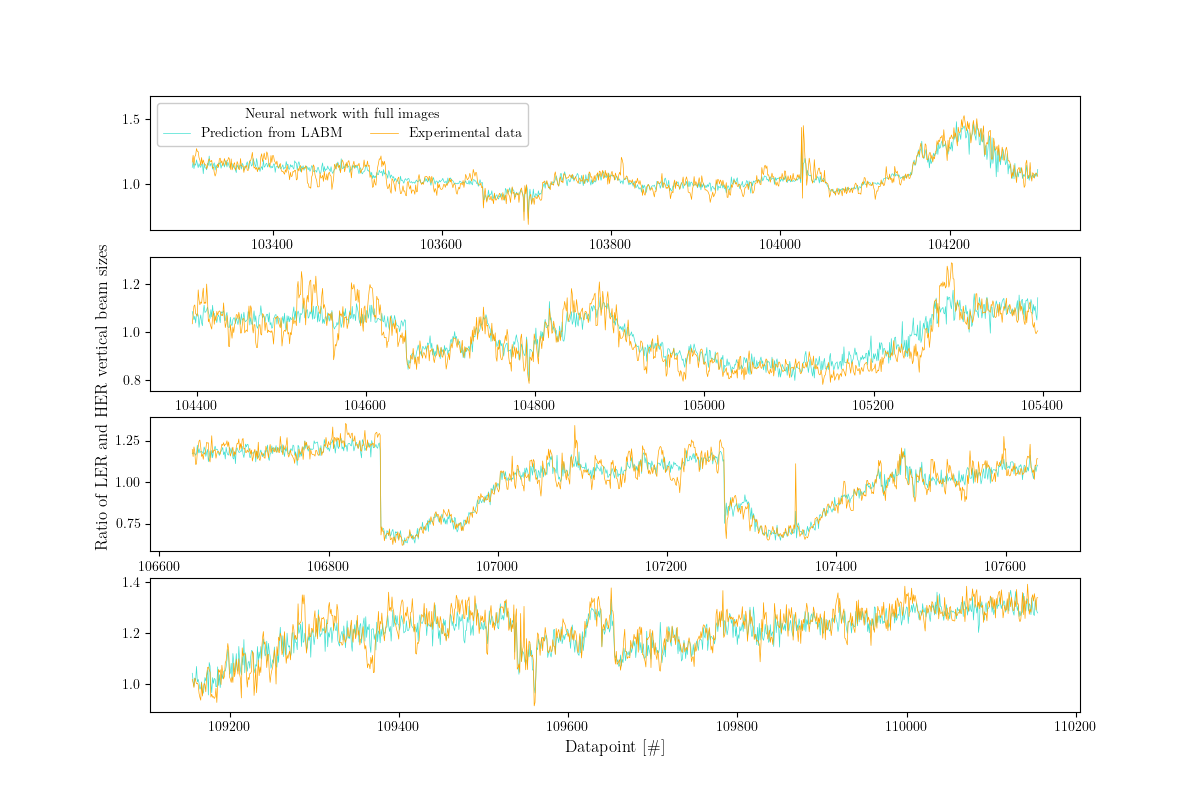}&
    \includegraphics[width=0.49\textwidth,trim=1.7cm 0.2cm 1.7cm 0.2cm,clip=true]{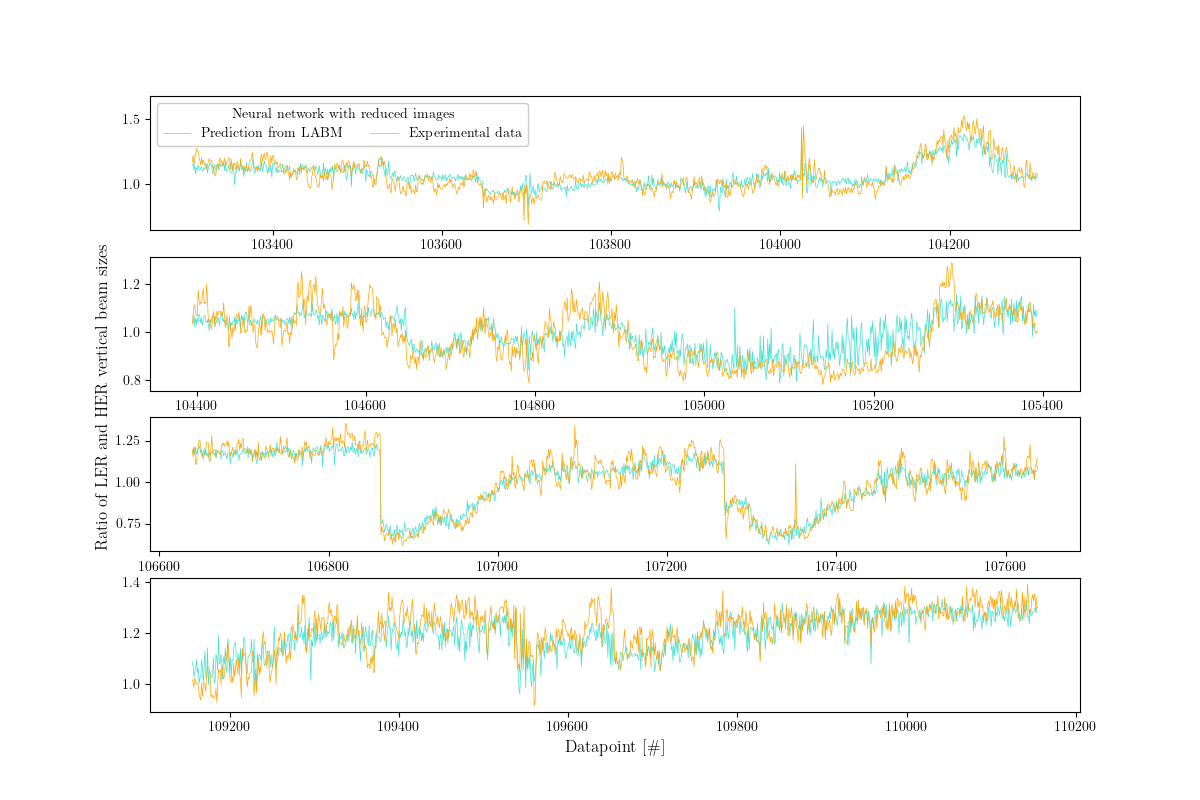}\\
  \end{tabular}
  \caption{Ratio of LER and HER vertical sizes
    $\sigma_{y,LER}/\sigma_{y,HER}/$ obtained with the upgraded setup
    with full images CNN \textit{(left)} and reduced images NN
    \textit{(right)}. Datapoints are sorted in chronological order.}
  \label{p:res-lhsigy-chrono}
\end{figure}

\begin{figure}[htp]
  \begin{tabular}{cc}
    \includegraphics[width=0.49\textwidth,trim=1.7cm 0.2cm 1.7cm 0.2cm,clip=true]{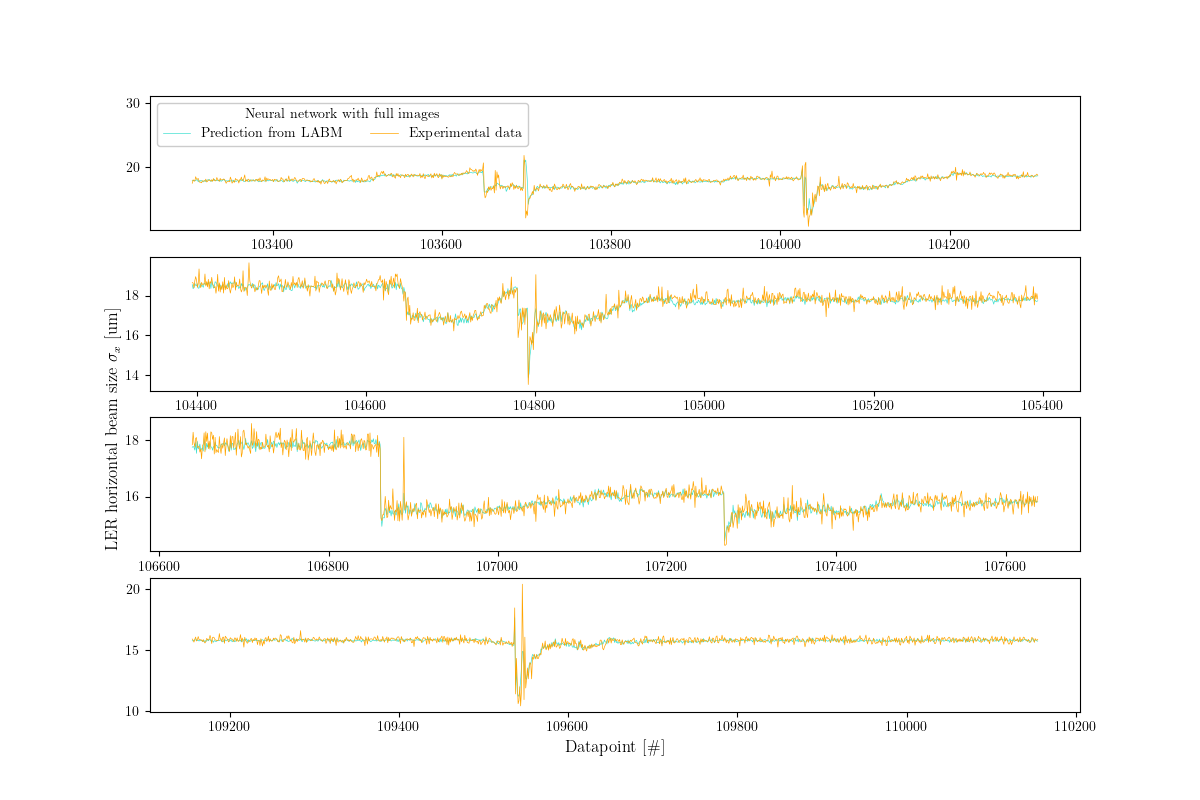}&
    \includegraphics[width=0.49\textwidth,trim=1.7cm 0.2cm 1.7cm 0.2cm,clip=true]{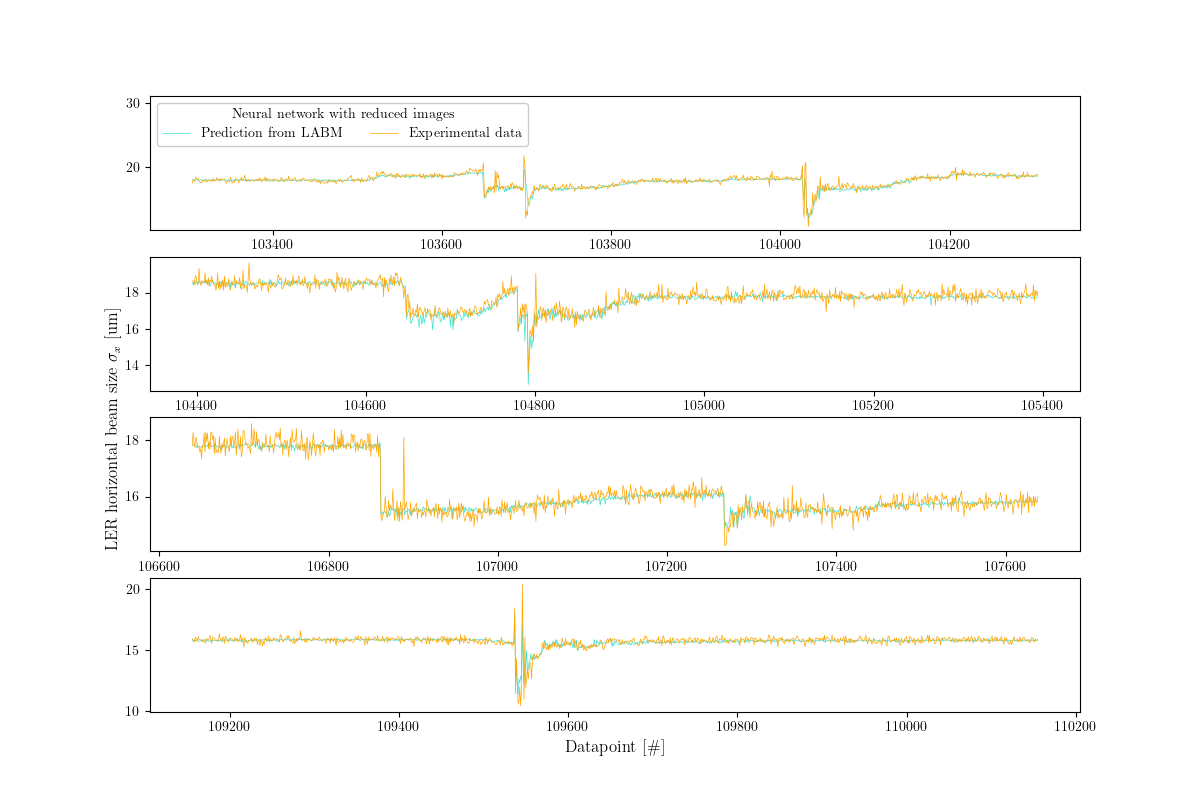}\\
  \end{tabular}
  \caption{LER horizontal size $\sigma_{x,LER}$ obtained with the upgraded
    setup with full images CNN \textit{(left)} and reduced images NN
    \textit{(right)}. Datapoints are sorted in chronological order.}
  \label{p:res-lsigx-chrono}
\end{figure}

\begin{figure}[htp]
  \begin{tabular}{cc}
    \includegraphics[width=0.49\textwidth,trim=1.7cm 0.2cm 1.7cm 0.2cm,clip=true]{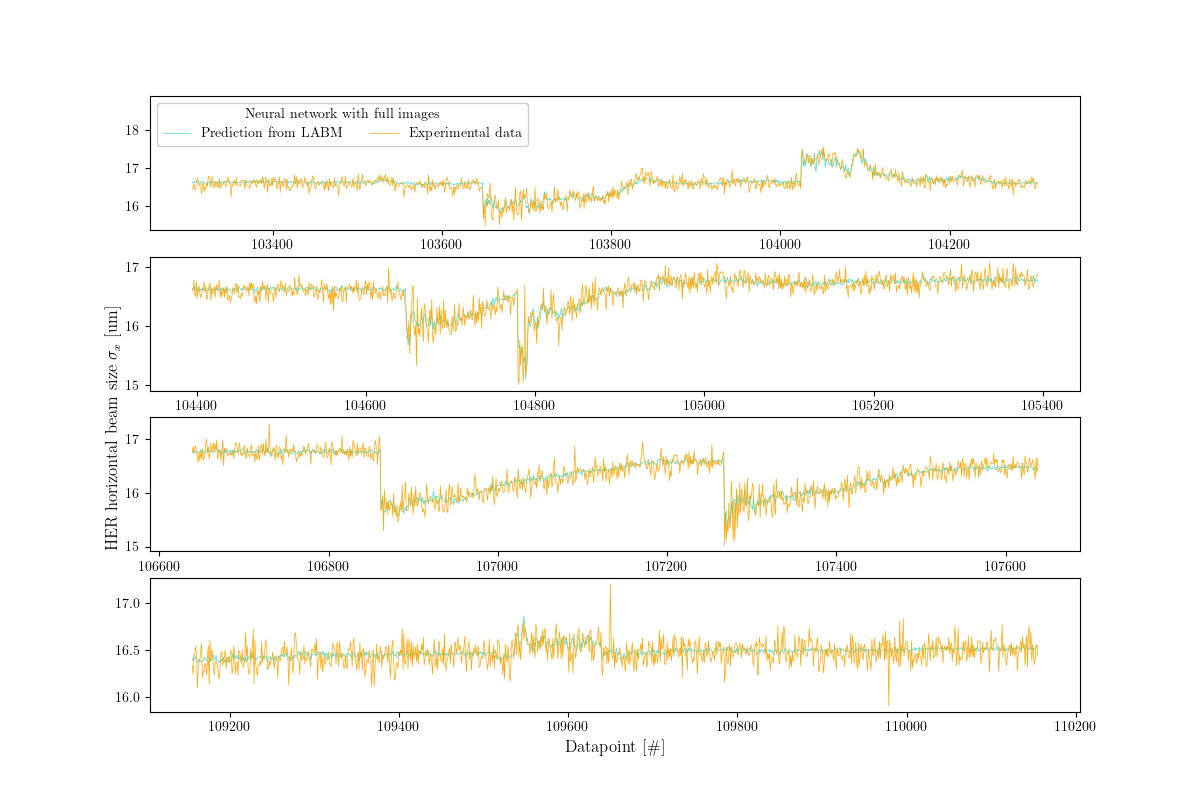}&
    \includegraphics[width=0.49\textwidth,trim=1.7cm 0.2cm 1.7cm 0.2cm,clip=true]{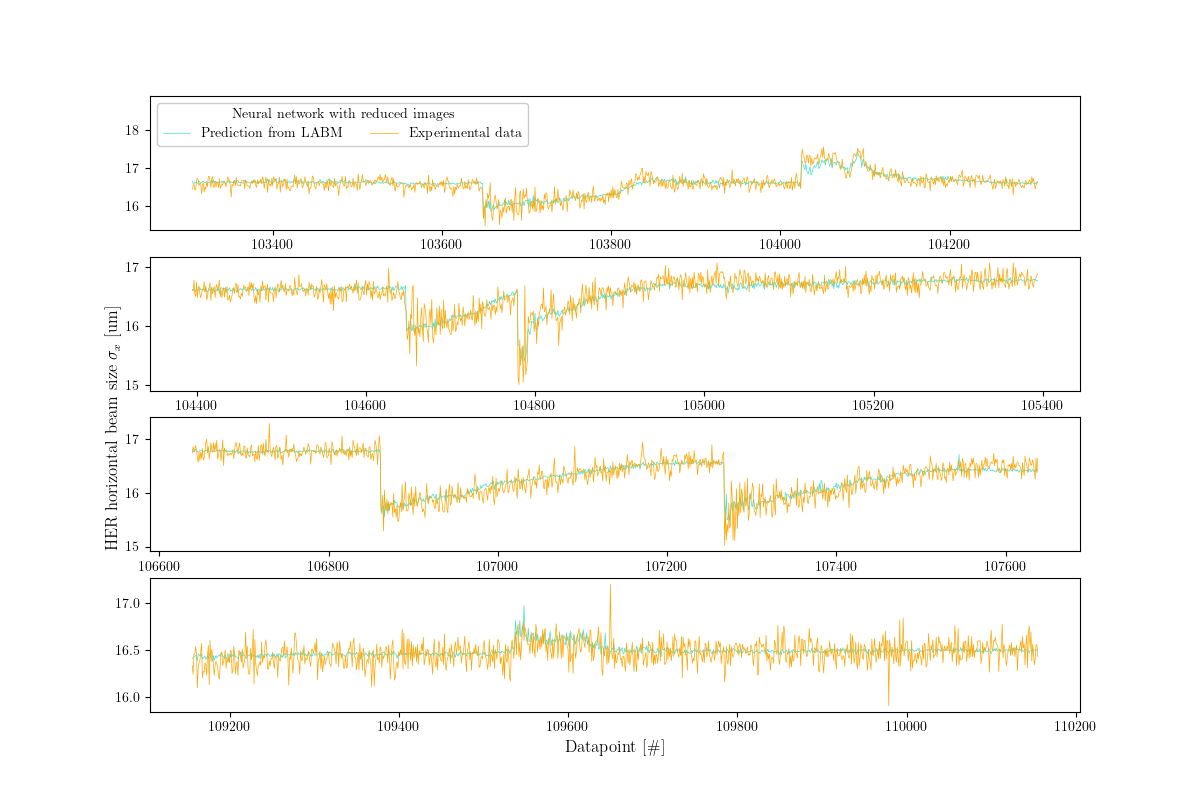}\\
  \end{tabular}
  \caption{HER horizontal size $\sigma_{x,HER}$ obtained with the upgraded
    setup with full images CNN \textit{(left)} and reduced images NN
    \textit{(right)}. Datapoints are sorted in chronological order.}
  \label{p:res-hsigx-chrono}
\end{figure}

\begin{figure}[htp]
  \begin{tabular}{cc}
    \includegraphics[width=0.48\textwidth,trim=1.7cm 0.2cm 1.7cm 0.2cm,clip=true]{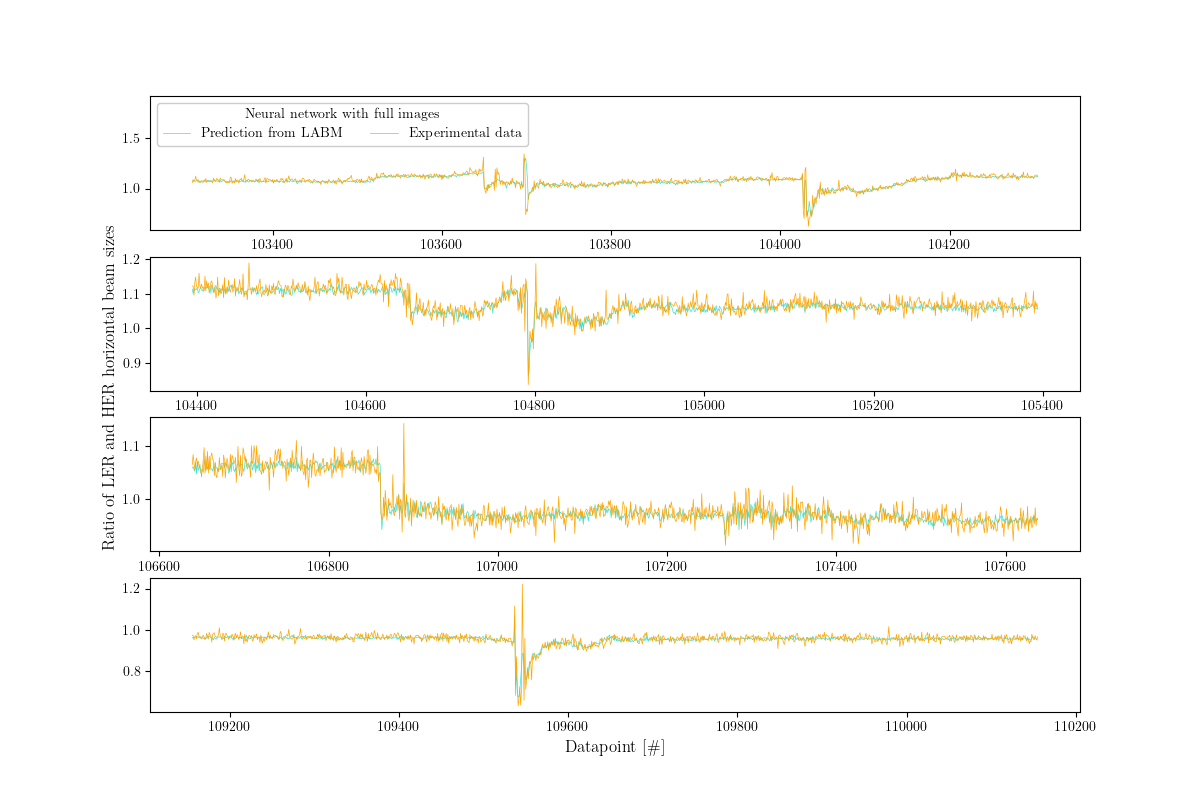}&
    \includegraphics[width=0.48\textwidth,trim=1.7cm 0.2cm 1.7cm 0.2cm,clip=true]{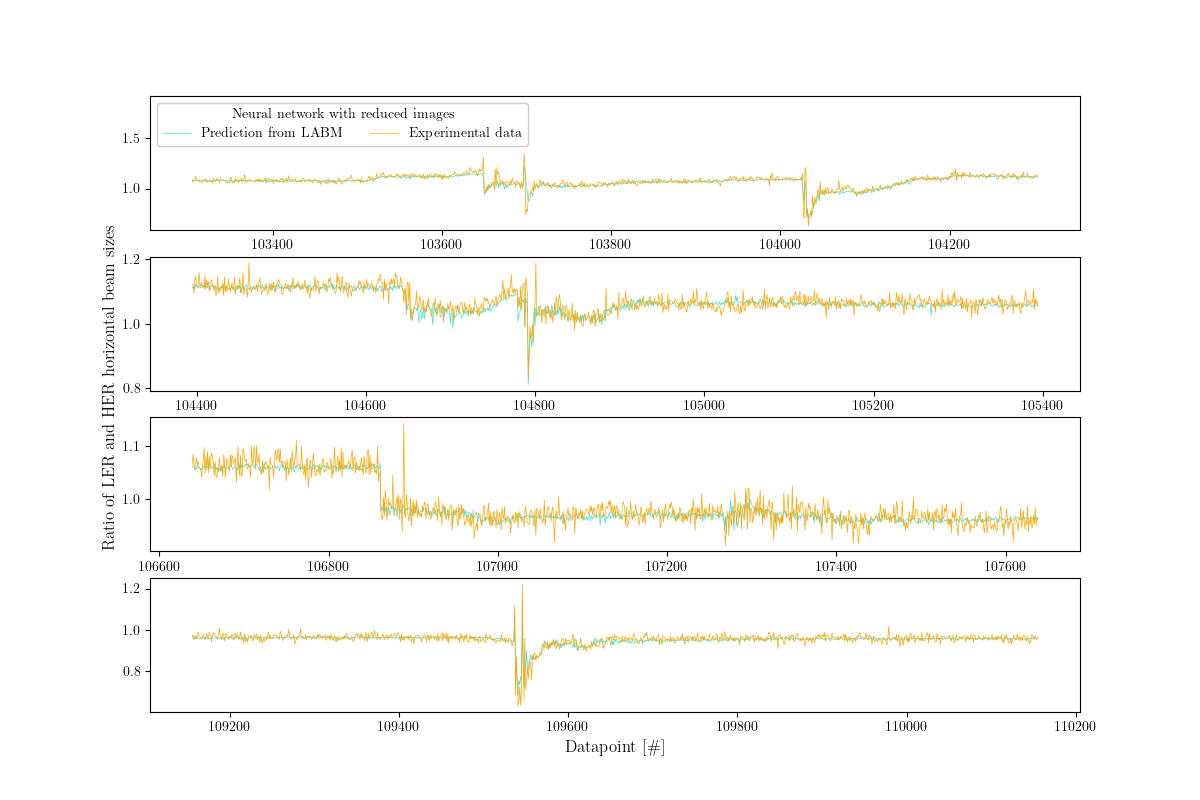}\\
  \end{tabular}
  \caption{Ratio of LER and HER horizontal sizes
    $\sigma_{x,LER}/\sigma_{x,HER}$ obtained with the upgraded setup
    with full images CNN \textit{(left)} and reduced images NN
    \textit{(right)}. Datapoints are sorted in chronological order.}
  \label{p:res-lhsigx-chrono}
\end{figure}

While the averaged predictions from our models follow the input
(measured) values well, individual points may significantly deviate from
them. To understand the reason for these discrepancies we plotted the
consecutive LABM predictions and the input measurements in chronological
order for four time windows from the test data subset, as shown in
Figs.~\ref{p:res-lumspec-chrono} to \ref{p:res-lhsigx-chrono}. Each
window is 1000 points long, which is 8000 seconds of measurements. We
see that the predictions follow the average of the measured values, but
the latter fluctuate much more. This may be explained by a relatively
long exposure time of the LABM cameras, $0.1\,\mathrm{s}$, compared to
$800\,\mathrm{\mu s}$ of the XRM cameras~\cite{xrm}. There are also
occasional short delays between significant changes in measured values
and LABM predictions. This may be due to poor synchronization between
LABM and the input measurements. Both shortening the exposure time and
better synchronization are in our TODO list for the future.


\begin{table}[!htbp]
\caption{Comparison of the results obtained with the previous setup
  \textit{(PMT)}~\cite{nnresult}, and with two different methods from
	this study \textit{(FI and RI)}, with neural networks
  \textit{(NN)} and linear regression \textit{(LR)}.}
\label{t:comp}
\begin{center}
\begin{tabular}{l|c|c|c|c|c|c}
  \hline
  \multirow{2}{*}{Variable} & \multicolumn{2}{c|}{PMT} & \multicolumn{2}{c|}{FI} & \multicolumn{2}{c}{RI} \\
  \cline{2-7}
                                  & NN     & LR     & NN     & LR      & NN      & LR \\
  \hline
  \hline
  $\mathcal{L}_{spec}$            & 3.4\%  & 4.2\%  & 1.50\% & 3.83\%  & 1.68\%  & 2.41\%  \\
  $\sigma_{y,e\!f\!f}$            & 2.2\%  & 3.7\%  & 1.24\% & 2.95\%  & 1.40\%  & 1.92\%  \\
  $\sigma_{y,LER}$                & 3.7\%  & 4.5\%  & 2.39\% & 5.68\%  & 2.74\%  & 3.97\%  \\
  $\sigma_{y,HER}$                & 3.4\%  & 6.6\%  & 2.41\% & 6.14\%  & 2.87\%  & 4.82\%  \\
  $\sigma_{y,LER}/\sigma_{y,HER}$ & 4.8\%  & 9.0\%  & 4.15\% & 10.27\% & 4.92\%  & 7.72\%  \\
  $\sigma_{x,LER}$                & 0.7\%  & 1.0\%  & 1.22\% & 2.44\%  & 1.25\%  & 1.47\%  \\
  $\sigma_{x,HER}$                & 0.9\%  & 1.0\%  & 0.64\% & 0.96\%  & 0.66\%  & 0.78\%  \\
  $\sigma_{x,LER}/\sigma_{x,HER}$ & 1.1\%  & 1.4\%  & 1.36\% & 2.44\%  & 1.40\%  & 1.66\%  \\
\end{tabular}
\end{center}
\end{table}

\section{Conclusion}

The primary goal of this publication was to study the implementation of
the cameras instead of PMTs as a signal detector for LABM. We could
successfully reproduce and improve our previous results by using much
less equipment, time and efforts and focus more on data analysis instead
of data acquisition.

\section{Future developments}

Using cameras instead of PMTs opens new possibilities for the LABM
development. The optical box may be miniaturized and moved next to the
beamline, in which case we greatly reduce number of mirrors in the
telescopes. It makes the device adjustment easier, number of
reflections and false spots smaller and the signal brighter. A possible
disadvantage of this change is the radiation level the device will
have to cope with. However our estimations show that the cameras
should be able to survive it for at least several years.

With a minimal exposure time of the cameras of order of $10\,\mathrm{\mu
  s}$, we can measure very fast beam changes, though we may be limited
by the amount of light and the sensor sensitivity.

Another issue is how we get the desirable results from the
beamstrahlung observation. At the moment LABM relies on the external
measurements from the accelerator group and Belle~II detector for the
training of the NNs. Another direction of the device development is to
make completely independent measurements of the beam parameters by
using beamstrahlung simulation. To realize this idea we communicate
with two groups working on beam simulation packages, SAD~\cite{sad}
and Xsuite~\cite{xsuite}. At the moment none of them provides complete
photon simulation but both may add it in the future.

EIC is the electron-ion collider at BNL. Even though only one beam is
electron and LABM usage for the proton beam is yet be studied, we
showed that even monitoring a single beam may give valuable
information. In future LABM-like device may be installed on
EIC~\cite{labm-eic}.

Another accelerator which may greatly benefit from the beamstrahlung
monitoring is FCC-ee~\cite{fccee}. The intensity of the beam and
therefore beamstrahlung brightness as well as radiation levels around
the accelerator make it challenging, but the benefits of the beam
monitoring through beamstrahlung are indisputable.


\end{document}